\documentclass[aps,prd,preprint,superscriptaddress,amsfonts,amssymb,amsmath,nofootinbib]{revtex4-1}

\usepackage{graphicx, bm, color, ulem, latexsym}
\usepackage[colorlinks,linkcolor=magenta,anchorcolor=blue,citecolor=blue]{hyperref}

\begin{document}
\tolerance=5000


\title{
Scalar-Induced Electromagnetic Radiation: Comparison with Axion-Like Particles and Implications for Modified Gravity
}


\author{Wenyi Wang}
\email{wangwy@mails.ccnu.edu.cn}
\affiliation{Institute of Astrophysics, Central China Normal University, Wuhan 430079, China}

\author{Sousuke Noda}
\email{snoda@cc.miyakonojo-nct.ac.jp}
\affiliation{National Institute of Technology, Miyakonojo College, Miyakonojo 885-8567, Japan}

\author{Taishi~Katsuragawa}
\email{taishi@ccnu.edu.cn}
\affiliation{Institute of Astrophysics, Central China Normal University, Wuhan 430079, China}


\begin{abstract}
The scalar-tensor theory of gravity, a modified gravity theory, introduces a fundamental scalar field that can serve as dynamical dark energy, driving the late-time accelerated expansion of the Universe. 
In this work, we analyze electromagnetic (EM) radiations arising from scalar fields and compare these features with those induced by axion-like particles (ALPs).
Scalar and ALP fields couple differently to the EM field due to their distinct parity properties, $\phi F_{\mu\nu} F^{\mu\nu}$ for scalar fields and $\phi F_{\mu\nu} \tilde{F}^{\mu\nu}$ for ALPs.
Building on analytical methods developed for ALPs, this work presents a theoretical feasibility analysis that demonstrates how the scalar field could produce observable EM signatures from oscillating field configurations.
We also show that resonance effects can amplify the EM radiation for the scalar field under specific conditions, and that the enhancement mechanisms depend on the coupling structure and the configuration of the background magnetic field. 
Resonance phenomena can accentuate the differences in signal strength and spectral features, potentially aiding future observations in distinguishing scalar fields from ALPs.
Although our studies apply to general scalar fields, we embed them within the framework of scalar-tensor theory and discuss the mass and coupling parameter in the context of testing modified gravity.
This work provides a theoretical framework for studying generic pure and pseudo-scalar fields on an equal footing and suggests new avenues for observational tests of modified gravity scenarios alongside ALP models.
\end{abstract}

\maketitle

\section{Introduction}
Axion and axion-like particles (ALPs) are among the most compelling candidates for physics beyond the Standard Model of particle physics~\cite{Peccei:1977hh, PhysRevLett.40.223, Cheng:1987gp}~\footnote{
The essential distinction between the axion and ALPs lies in their origin and purpose. 
While the axion is introduced to resolve the strong CP problem in quantum chromodynamics, ALPs are pseudoscalar particles that arise in broader theoretical frameworks such as string theory.
}. 
From the perspectives of both particle physics and cosmology, various types of ALPs have been extensively studied in the contexts of inflation~\cite{Silverstein:2008sg, Pajer:2013fsa, Freese:2014nla}, dark matter (DM)~\cite{Preskill:1982cy, Abbott:1982af, Dine:1982ah, Kim:1986ax, Duffy:2009ig, Arvanitaki:2009fg, Marsh:2015xka}, and dark energy (DE)~\cite{Kim:2002tq, Chacko:2004ky}.
In contrast to axion and ALPs, modified gravity theories predict the pure scalar fields~\cite{Sotiriou:2008rp}.
These scalar fields arise as additional degrees of freedom in scalar–tensor theories and $F(R)$ gravity theories, and they can be interpreted as the inflaton~\cite{Linde:1981mu, Linde:1983gd}, dynamical DE~\cite{Ratra:1987rm, Chen:2022zkc}, and even DM candidate~\cite{Nojiri:2008nk, Nojiri:2008nt, Cembranos:2008gj, Choudhury:2015zlc, Katsuragawa:2016yir, Burrage:2016yjm, Burrage:2018zuj, Chen:2019kcu, Shtanov:2021uif, Kading:2023hdb, Shtanov:2024nmf}. 
While the scalar fields share similar physical applications with ALPs, their theoretical motivations differ significantly.

Axion and ALPs, being pseudoscalar fields, couple to the EM field through the $F_{\mu\nu}\widetilde{F}^{\mu\nu}$ operator, whereas scalar fields can couple via the $F_{\mu\nu} F^{\mu\nu}$ operator, which is associated with the trace anomaly
~\cite{Fujikawa:1980vr, Fujii:2015iuz, Ferreira:2016kxi}.
Their couplings to the EM field may result in distinct features in the emitted EM radiation. 
Many ongoing and planned experiments aim to detect these new fields via their coupling to the electromagnetic (EM) field~\cite{ADMX:2010ndb, XENON:2020rca, Homma:2019rqb, Salemi:2021gck, Chou:2008gr, Steffen:2010ze, Rybka:2010ah, Brax:2009aw, CAST:2018bce, Vagnozzi:2021quy, Katsuragawa:2021wmw}.
Given the differences in their coupling forms, it is instructive to investigate how methods and setups developed for ALP searches can be applied to pure scalar fields, potentially revealing complementary observational signatures.

Since the couplings of these fields, the pure scalars and ALPs, to EM fields are generally weak, detection often requires strong EM backgrounds or large field amplitudes.
Astrophysical environments such as rotating neutron stars, compact binaries, and neutron star mergers offer promising conditions with strong EM fields that could enhance detectability.
Dense axion stars have been investigated in astrophysical contexts; for instance, radiation from dense axion star-neutron star binaries~\cite{Kyriazis:2022gvw}, axion stars as candidates for planet 9~\cite{Di:2023xaw}, and axion condensate around a black hole~\cite{Yoshino:2013ofa, Omiya:2022gwu}. 
Furthermore, if the pure scalar fields or ALP fields exhibit coherent oscillations, resonance effects can significantly amplify otherwise weak couplings, leading to detectable EM radiation.
As shown in Refs.~\cite{Amin:2021tnq, Sen:2021mhf, Redondo:2008ec, Redondo:2013lna, Hardy:2024gwy}, such resonance enhancement occurs for ALPs when the plasma frequency in the surrounding medium approximates the ALP mass scale.
Similar effects can arise when the frequency of an alternating magnetic field matches the ALP mass~\cite{Sen:2021mhf}.

In this work, building on the theoretical framework and analytic methodology employed in ALP studies, we explore the potential observability of pure scalar fields via their EM couplings.
In particular, this study focuses on oscillating spherical field configurations in the presence of a background magnetic field.
Applying special-relativistic calculation methods developed to study EM emissions from ALPs condensates, we evaluate the strength of EM radiation produced by the scalar under several background EM field configurations. 
Finally, we compare the detectability of EM signals produced by pure scalar fields with that of signals generated by the ALPs.

Compared with the results in the ALP case, analytic calculations show that the scalar field predicts qualitative differences in the basic equations induced by distinct coupling structures to the EM field.
Numerical calculations demonstrate that the scalar field can generate observable EM signals in a specific region of mass and coupling, as the existing works on axion and ALPs have predicted observable EM signals.
Applying our considerations to the scalar-tensor theory as a benchmark scenario, we find that the parameter space of the scalar-field mass and coupling overlaps with that of existing and planned ground-based experiments, potentially allowing us to constrain modified gravity theories synergetically.
While scalar and ALP fields may yield comparable EM radiation power, distinct resonance behaviors could provide an additional means to distinguish between them in future observations.

This paper is organized as follows. 
In Sec.~2, we briefly introduce the scalar-tensor theory and discuss the scalar field coupling to the EM field.
In Sec.~3, we introduce the pure and pseudo-scalar fields coupled to the EM field and derive the corresponding field equations. 
Using the perturbative approach, we then formulate the EM radiation produced by the oscillating scalar and ALP fields.
In Sec.~4 and Sec.~5, we examine the EM radiation power generated under different background field configurations.
In Sec.~6, we analyze the qualitative behavior of the radiation power, 
highlighting the differences and similarities between the two cases. 
We also assess the detectability of EM signals generated by the scalar and ALP.
In Sec.~7, we apply our results in Sec.~4 and Sec.~6 within the framework of scalar-tensor theory and discuss the parameter regions for mass and coupling constant.
Finally, Sec.~8 is devoted to the conclusions and discussion of our results.   
Throughout this paper, we use the natural unit system: $G = \hbar = c =1$.

\section{Scalar-Tensor theory}

The scalar-tensor theory has been investigated in cosmology to account for the DE driving the late-time accelerated expansion of the Universe~\cite{Ratra:1987rm, Nojiri:2017ncd, Clifton:2011jh, Chen:2022zkc}.
In this section, we briefly introduce the scalar-tensor theory and the emergence of the coupling between the scalar and EM fields.

\subsection{Coupling between scalar field and EM field}

The simplest model for the scalar-tensor theory is general relativity with a minimally coupled scalar field:
\begin{align}
\label{Eq: ST theory}
\begin{split}
    S 
    &= 
    \int\!\mathrm{d^4}x\,\sqrt{-g} 
    \left[
        \frac{1}{2\kappa^2} R - \frac{1}{2} g^{\mu\nu} \partial_{\mu}\phi \partial_{\nu}\phi -V(\phi) 
    \right]
    + S_{M} [A^2(\phi) g_{\mu\nu}, \psi]
    \, .
\end{split}
\end{align}
Here, $1/\kappa = M_{\text{pl}}$ is the (reduced) Planck mass, and $S_{M}$ represents the action for the matter field $\psi$, such as the EM field.
$V(\phi)$ is a potential of the scalar field, including the mass term and self-interactions, and $A(\phi)$ represents the coupling to the matter field $\psi$.

The coupling between the scalar field $\phi$ and matter field $\psi$ shows up from the spacetime metric, and the interaction is written as a function $A(\phi)$.
Considering the Taylor expansion of the coupling function $A(\phi)$ and the interaction between the scalar field and EM field, we find that the scalar-tensor theory naturally realizes our consideration:
\begin{align}
\begin{split}
    \mathcal{L}
    &=
    - \frac{1}{4} A^4 (\phi) F_{\mu\nu} F^{\mu\nu}
    \\
    &\approx
    - \frac{1}{4} F_{\mu\nu} F^{\mu\nu}
    - \frac{g_{s\gamma}}{4} \phi F_{\mu\nu} F^{\mu\nu}
    \, .
\end{split}
\end{align}
In general scalar-tensor theory, $g_{s\gamma}$ can take arbitrary value~\cite{brans1961mach,horndeski1974second,fujii2003scalar,faraoni2004scalar}.
Note that this coupling can be interpreted as the trace anomaly induced by the fermion loop diagrams~\cite{Fujikawa:1980vr, Fujii:2015iuz, Ferreira:2016kxi}. 
In the scalar-tensor theory, when multiple fermions are involved, the coupling constants between the scalar field and each fermion can be chosen independently; consequently, the coupling constant $g_{s\gamma}$, which incorporates contributions from all fermions, can be parametrized.

\subsection{Chameleon mechanism and effective mass}

For the DE problem, the scalar field $\phi$ plays a role as the dynamical DE, and its potential $V(\phi)$ should involve the DE scale.
However, the scalar-field mass can be much larger than the DE scale due to the chameleon mechanism~\cite{Khoury:2003aq, Khoury:2003rn} triggered by the ambient matter contents.
Variation of Eq.~\eqref{Eq: ST theory} with respect to the scalar field yields the following Klein-Gordon equation:
\begin{align}
\begin{split}
    0
    &=
    \Box \phi - V^{\prime}(\phi) + \beta(\phi) A^{4}(\phi) T^{\mu}_{\ \mu} 
    \, , \\
    \beta(\phi)
    & \equiv
    \left( \ln A(\phi) \right)^{\prime}
    \, ,
\end{split} 
\end{align}
where $T^{\mu}_{\ \mu}$ is the trace of the energy-momentum tensor $T_{\mu\nu}$ of matter field $\psi$ and metric $g_{\mu\nu}$.
Defining the effective potential $V_\mathrm{eff}$ as
\begin{align}
    V^{\prime}_\mathrm{eff}(\phi) 
    = 
    V^{\prime}(\phi) - \beta(\phi) A^{4}(\phi) T^{\mu}_{\ \mu}
    \, ,
\end{align}
we can evaluate the effective mass of the scalar field as
\begin{align}
    m^{2}_\mathrm{eff} 
    = 
    V^{\prime \prime}_\mathrm{eff}(\phi_{\min}) 
    \, ,
\end{align}
where $\phi_{\min}$ is given as $V^{\prime}(\phi_{\min})=0$.

By tuning the potential $V(\phi)$ and coupling function $A(\phi)$, it is possible to realize that the effective mass $m_\mathrm{eff} $ is an increasing function of $-T^{\mu}_{\ \mu}$, and this is called the chameleon mechanism.
For example, in the simple case that $-T^{\mu}_{\ \mu}=\rho$ for the non-relativistic fluid with the energy density $\rho$, we find that the chameleon mechanism makes the scalar field heavy in the high-density environment~\footnote{
This is the effective mass defined in the interacting system and can be understood as a medium effect.
}.

For instance, the simplest mass term $V(\phi)=\frac{1}{2} m^2_s \phi^2$ can be embedded into the potential $V(\phi)$, where $m_s$ is the bare mass of the scalar field.
In the DE model of scalar-tensor theory, the scalar-field mass is generally expected to be at the DE scale; however, the mass range of our interest lies in the heavier regime.
As shown later, it is possible to realize our targeted mass region within the framework of the scalar-tensor theory with the chameleon mechanism~\cite{Brax:2013ida, Burrage:2017qrf, Brax:2021wcv}.
We note that the effective mass depends on the model of the scalar-tensor theory and the ambient matter field.
In the following, to develop an approach for the general scalar field, we treat the effective mass as a mass parameter, as well as the coupling constant to the EM field.

\section{Model and perturbative approach to EM radiation}

In this section, we formulate the basic field equation describing an interacting system comprising the EM field, a pure/pseudo-scalar field, and a matter source.
Deriving the Klein-Gordon equation and Maxwell equation in the Minkowski spacetime, we observe the different coupling between the scalar/ALP and EM fields in terms of the electric and magnetic fields.
Then, we introduce several models of background oscillating scalar/ALP fields that have been used in existing works on EM radiation from ALPs.
We finally consider a perturbative approach to the EM radiation generated by the oscillation of scalar/ALP fields.

\subsection{Pure and pseudo scalars coupled to EM field}

We consider the following Lagrangian: 
\begin{align}
\label{Eq: lagrangian}
\begin{split}
    \mathcal{L}
    &= 
    - \frac{1}{2}g^{\mu\nu} (\partial_{\mu}\phi)(\partial_{\nu}\phi) 
    - V(\phi) 
    - \frac{1}{4}F_{\mu\nu}F^{\mu\nu} 
    \\
    & \qquad 
    + J_{m}^{\mu} A_{\mu} 
    - \frac{g_{s \gamma }}{4}\phi F_{\mu\nu} F^{\mu\nu} 
    - \frac{g_{a \gamma }}{4}\phi F_{\mu\nu}\widetilde{F}^{\mu\nu}
    \, .
\end{split}
\end{align}
$\phi(x,t)$ represents the pure or pseudo-scalar field coupled to the EM field $A_{\mu}(x)$, and $V(\phi)$ is the potential. 
$g_{s \gamma}$ and $g_{a \gamma}$ are the coupling constants in the case of pure scalar and pseudo scalar, respectively. 
$F_{\mu\nu}$ and $\widetilde{F}_{\mu\nu}$ are the EM field strength tensor and its dual, defined in terms of the EM field as
\begin{align}
    F_{\mu\nu}
    &=
    \partial_{\mu}A_{\nu}-\partial_{\nu}A_{\mu}
    \, , \\
    \widetilde{F}^{\mu\nu}
    &=
    \frac{1}{2}\epsilon^{\mu \nu \rho \sigma}F_{\rho \sigma}
    \, ,
\end{align}
where the Levi-Civita tensor is defined as $\epsilon^{0 1 2 3} =1$. 
$J^{\mu}_{m}$ is a matter current other than the scalar field sourcing the EM field $A_{\mu}$.

By setting $g_{a\gamma}=0$, 
the field equations with respect to the pure scalar field $\phi$ and the EM field $A_{\mu}$ are given as follows:
\begin{align}
\label{Eq: eom_scalar}
    0
    &=
    \Box \phi - V^{\prime}(\phi) - \frac{g_{s\gamma}}{4} F_{\mu\nu}F^{\mu\nu}
    \, , \\
\label{Eq: eom_elemag_scalar}
    0
    &=
    (1 + g_{s\gamma} \phi) \partial_{\mu} F^{\mu\nu}
    + J_{m}^{\nu}
    + g_{s\gamma} (\partial_{\mu} \phi) F^{\mu\nu}
    \, .
\end{align}
In Eq.~\eqref{Eq: eom_elemag_scalar}, we define the $4$-current arising from the interaction between EM and pure scalar fields as
\begin{align}
\label{Eq: 4current_elemag_scalar}
    J_{s}^{\nu} 
    = 
    \frac{g_{s\gamma}}{1 + g_{s\gamma}\phi} (\partial_{\mu} \phi) F^{\mu\nu}
    \, .
\end{align}
Eqs.~\eqref{Eq: eom_scalar} and \eqref{Eq: eom_elemag_scalar} are reduced to the following forms:
\begin{align}
\label{Eq: eom_scalar2}
    \Box \phi - V^{\prime}(\phi)
    &=
    \frac{g_{s\gamma}}{4} F_{\mu\nu} F^{\mu\nu}
    \, , \\
\label{Eq: eom_elemag_scalar2}
    \partial_{\mu} F^{\mu\nu}
    &=
    - \frac{1}{1 + g_{s\gamma} \phi} J_{m}^{\nu} - J_{s}^\nu
    \, .
\end{align}

In the same manner, by setting $g_{s\gamma}=0$, 
the field equation with respect to the pseudo scalar field $\phi$ and EM field $A_{\mu}$ are given as follows:
\begin{align}
\label{Eq: eom_axion}
    0
    &=
    \Box \phi - V^{\prime}(\phi) - \frac{g_{a\gamma}}{4} F_{\mu\nu} \widetilde{F}^{\mu\nu}
    \, , \\
\label{Eq: eom_elemag_axion}
    0
    &=
    \partial_{\mu} F^{\mu\nu}
    + J_{m}^{\nu}
    + g_{a\gamma} (\partial_{\mu} \phi) \widetilde{F}^{\mu\nu}
    + g_{a\gamma} \phi \left( \partial_{\mu} \widetilde{F}^{\mu\nu} \right)
    \, .   
\end{align}
Here, we use an identity for the dual of the EM field strength
\begin{align}
    \partial_{\mu} \widetilde{F}^{\mu\nu}=0
    \, 
\end{align}
and define the $4$-current arising from the interaction term as
\begin{align}
\label{Eq: 4current_elemag_axion}
    J_{s}^{\nu} = g_{a\gamma} (\partial_{\mu} \phi) \widetilde{F}^{\mu\nu}
    \, .
\end{align}
Then, Eqs.~\eqref{Eq: eom_axion} and \eqref{Eq: eom_elemag_axion} are reduced to 
\begin{align}
\label{Eq: eom_axion2}
    \Box \phi - V^{\prime}(\phi)
    &=
    \frac{g_{a\gamma}}{4} F_{\mu\nu} \widetilde{F}^{\mu\nu}
    \, , \\
\label{Eq: eom_elemag_axion2}
    \partial_{\mu} F^{\mu\nu}
    &=
    - J_{m}^{\nu}
    - J_{s}^\nu
    \, .
\end{align}

The pure scalar and pseudo scalar fields have different coupling to the EM field, $F_{\mu\nu}F^{\mu\nu}$ or $F_{\mu\nu}\widetilde{F}^{\mu\nu}$, as in Eqs.~\eqref{Eq: eom_scalar2} and \eqref{Eq: eom_axion2}.
In both cases, the current $J_s$ is proportional to the coupling constant $g_{s\gamma}$ or $g_{a\gamma}$.
However, different interactions with the EM field enable us to distinguish between the pure and pseudo-scalar fields, regardless of the value of the coupling constant.

\subsection{Klein-Gordon and Maxwell equations in flat spacetime}

We work in the flat spacetime $g_{\mu\nu} = \mathrm{diag}(-1, 1, 1, 1)$.
To write each component of the field equations, we define the electric and magnetic fields,
\begin{align}
\label{Eq: notations}
\begin{split}
    E_{i}
    &= F_{i0} 
    \, , \\
    B_{i}
    &= \frac{1}{2} \epsilon_{ijk} F^{jk}
    \, , 
\end{split}
\end{align}
where $\epsilon_{ijk} = \epsilon^{ijk}$ and $i,j,k = 1,2,3$.
We then write $F_{\mu\nu}F^{\mu\nu}$ and $F_{\mu\nu}\widetilde{F}^{\mu\nu}$ in terms of the electric field $\mathbf{E}$ and magnetic field $\mathbf{B}$:
\begin{align}
\label{Eq: fmunufmunu}
    F_{\mu\nu} F^{\mu\nu} 
    &= 
    - 2 \left( \mathbf{E}^2 - \mathbf{B}^2 \right)
    \, , \\
\label{Eq: fmunutilde}
    F_{\mu\nu}\widetilde{F}^{\mu\nu} 
    &=
    - 4 \mathbf{E} \cdot \mathbf{B}
    \, .
\end{align}
Moreover, we denote the $4$-currents $J_{s}^{\mu}$ and $J_{m}^{\mu}$ by
\begin{align}
\label{Eq: scalar current}
    J_{s}^{\mu} 
    &= 
    \left(\rho_{s}, \mathbf{J}_{s} \right)
    \, , \\
\label{Eq: matter current}
    J_{m}^{\mu} 
    &= 
    \left(\rho_{m}, \mathbf{J}_{m} \right)
    \, .
\end{align} 

In terms of the electric and magnetic fields, we derive the field equations for the pure and pseudo-scalar fields, as well as the EM field.
In the pure scalar case, Eq.~\eqref{Eq: eom_scalar2} leads to the Klein-Gordon equation sourced by the EM field,
\begin{align}
\label{Eq: kg_scalar}
    \ddot{\phi} - \nabla^{2} \phi + V^{\prime} 
    &= 
    \frac{g_{s \gamma}}{2} \left( \mathbf{E}^2 - \mathbf{B}^2 \right)
    \, , 
\end{align}
and Eq.~\eqref{Eq: eom_elemag_scalar2} leads to the Maxwell equation sourced by the pure scalar field and the other matter,
\begin{align}
\label{Eq: maxwell_scalar_pert}
\begin{split}
    \nabla \times \mathbf{B} - \dot{\mathbf{E}} 
    &= 
    \frac{1}{1 + g_{s \gamma} \phi} \mathbf{J}_{m} + \mathbf{J}_{s}
    \, ,\\
    \nabla \times \mathbf{E} + \dot{\mathbf{B}}
    &= 
    0
    \, , \\
    \nabla \cdot \mathbf{B} 
    &= 0
    \, , \\
    \nabla \cdot \mathbf{E} 
    &= 
    \frac{1}{1 + g_{s \gamma} \phi} \rho_{m} + \rho_{s}
    \, .
\end{split}
\end{align}
From Eq.~\eqref{Eq: 4current_elemag_scalar},
the charge density $\rho_s$ and current vector ${\bf J}_s$ are written as
\begin{align}
\label{Eq: charge_elemag_scalar}
    \rho_{s}
    &= 
    - \frac{g_{s \gamma}}{1 + g_{s \gamma} \phi} 
    \nabla \phi \cdot \mathbf{E} 
    \, , \\
\label{Eq: 3current_elemag_scalar}
    {\bf J}_{s}
    &= 
    \frac{g_{s \gamma}}{1 + g_{s \gamma} \phi} 
    \left(\dot{\phi} \mathbf{E} - \nabla \phi \times \mathbf{B} \right)
    \, .
\end{align}

In the pseudoscalar case, 
Eq.~\eqref{Eq: eom_axion2} leads to the Klein-Gordon equation sourced by the EM field,
\begin{align}
\label{Eq: kg_axion}
    \ddot{\phi}-\nabla^{2} \phi + V^{\prime} 
    &= 
    g_{a \gamma} \mathbf{E} \cdot \mathbf{B}
    \, , 
\end{align}
and Eq.~\eqref{Eq: eom_elemag_axion2} leads to the Maxwell equation sourced by the pseudoscalar field and the other matter,
\begin{align}
\label{Eq: maxwell_axion_pert}
\begin{split}
    \nabla \times \mathbf{B} - \dot{\mathbf{E}} 
    &= 
    {\bf J}_{m}
    + {\bf J}_{s} 
    \, , \\
    \nabla \times \mathbf{E} + \dot{\mathbf{B}}
    &= 0
    \, ,\\
    \nabla \cdot \mathbf{B} 
    &= 0
    \, ,\\
    \nabla \cdot \mathbf{E} 
    &= 
    \rho_m + \rho_{s} 
    \, ,
\end{split}
\end{align}
where
\begin{align}
\label{Eq: charge_elemag_axion}
    \rho_{s} 
    &= 
    -g_{a \gamma } \nabla \phi \cdot \mathbf{B}
    \, , \\
\label{Eq: 3current_elemag_axion}
    {\bf J}_{s} 
    &= 
    g_{a \gamma }\left(\dot{\phi} \mathbf{B} + \nabla \phi \times \mathbf{E} \right)
    \, .
\end{align}

\subsection{EM radiation from spherical scalar/ALP field condensate}

Hereafter, we consider the ALP as the pseudoscalar field and use the term \textit{scalar} for the pure scalar field.
To analyze the characteristics and differences of the EM radiation generated by the scalar and ALP fields, we assume a spherically symmetric, oscillating field configuration of the following form~\cite{Amin:2021tnq, Sen:2021mhf, Seidel:1991zh}:
\begin{align}
\label{Eq: bkgscalar1}
    \phi \left (\mathbf{x}, t\right) 
    = 
    \phi_{0}~{\rm sech}
    \left( \frac{\left|\mathbf{x} \right|}{R} \right)
    \cos \left( \omega t \right)
    ~,
\end{align}
where $\mathbf{x}:=(x,y,z)$, 
$\phi_{0}$ is a constant representing the amplitude and $\omega$ is the frequency of the time-varying field, and $R$ is the typical size of the field configuration. 
In this work, we mainly consider $\omega \sim m $, where $m$ is the scalar or ALP field mass.
In addition to the above setup proposed in existing works, we also consider a variant that describes the oscillating field configuration with a time-dependent $R$, $R(t)$:
\begin{align}
\label{Eq: bkgscalar2}
\begin{split}
    \phi \left (\mathbf{x}, t\right) 
    &= 
    \phi_{0}~{\rm sech} 
    \left[ \frac{\left|\mathbf{x} \right|}{R(t)} \right]
    \, , \\
    R(t) 
    &= 
    R \left[ 1 + \delta_R \cos (\omega_{o} t) \right]
    \, ,
\end{split}
\end{align}
where $\delta_{R} \ll 1$, and $\omega_{o}$ is the frequency of radius oscillation.
Taylor expansion with respect to a small $\delta_R$ leads to a configuration similar to the time-independent part of Eq.~\eqref{Eq: bkgscalar1} at the leading order.
The above oscillation reflects the existing study of a radially oscillating (breathing) scalar field configuration~\cite{Gleiser:1988ih, Kain:2021rmk}.

\begin{figure}[htbp]
    \begin{minipage}[b]{0.48\columnwidth}
        \centering
        \includegraphics[width=\columnwidth]{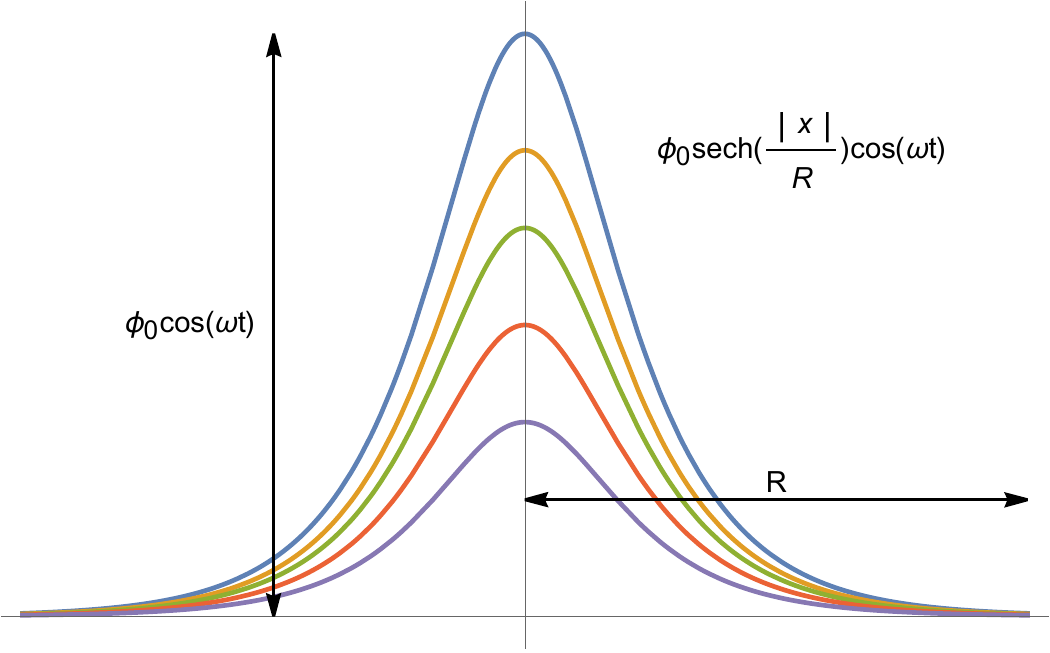}
    \end{minipage}
    \begin{minipage}[b]{0.48\columnwidth}
    \centering
    \includegraphics[width=\columnwidth]{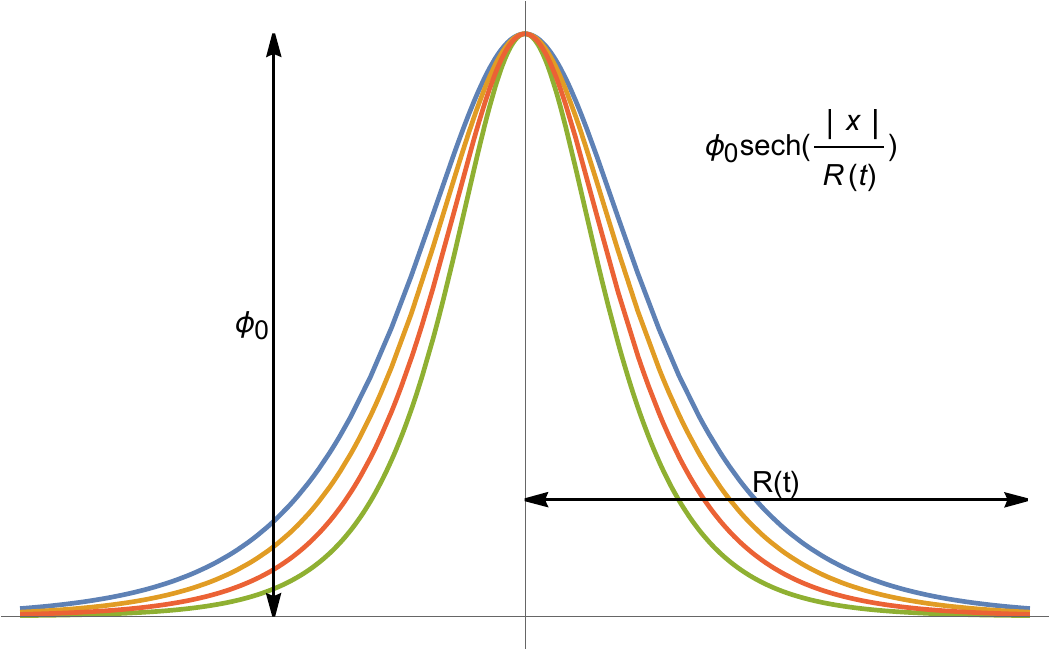}
    \end{minipage}
    \caption{
    Plots of the spherically symmetric, oscillating field configuration $\phi \left(\mathbf{x},t\right)$ (Left panel) and the oscillating field configuration $\phi \left(\mathbf{x},t\right)$ with a time-dependent $R(t)$ for $\delta_R = 1/5$ (Right panel). 
    Different colors represent the distribution of the field at different moments. 
    } 
    \label{schematicPhi}
\end{figure}

We note that Eqs.~\eqref{Eq: bkgscalar1} and \eqref{Eq: bkgscalar2} are ansatz in our analysis, which do not represent a fully self-consistent solution of the coupled Einstein–scalar/ALP field equations. 
Our study aims to develop a theoretical analysis of the scalar and ALP fields in the same manner and to capture the essential features and differences in their couplings to the EM field.
Thus, instead of fully solving the field equations with concrete settings, we analyze EM radiation from the spherically symmetric ansatz using the established methodology for astrophysical searches for the axion and ALPs.
We plot Eqs.~\eqref{Eq: bkgscalar1} and \eqref{Eq: bkgscalar2} in Fig.~\ref{schematicPhi}.

To study the radiated EM field, we consider an analytic treatment in the limit of a small coupling constant and expand the EM field as follows
\begin{align}
    \mathbf{E} \left(\mathbf{x},t\right) 
    &=
    \mathbf{E}_{0}\left(\mathbf{x},t\right) 
    + \mathbf{E}_{r} \left(\mathbf{x},t\right) 
    \, , \\ 
    \mathbf{B} \left(\mathbf{x},t\right) 
    &=
    \mathbf{B}_{0}\left(\mathbf{x},t\right) 
    + \mathbf{B}_{r} \left(\mathbf{x},t\right) 
    \, .
\end{align}
$\mathbf{E}_r, \mathbf{B}_r$ represent the perturbed (radiated) EM field from the background $\mathbf{E}_{0}, \mathbf{B}_{0}$,
and $\mathbf{E}_r, \mathbf{B}_r \ll \mathbf{E}_{0}, \mathbf{B}_{0}$.
The current $J_{s}^{\mu}$ in Eqs.~\eqref{Eq: 4current_elemag_scalar} and \eqref{Eq: 4current_elemag_axion} is proportional to the coupling constant, and we write the matter current as 
\begin{align}
\begin{split}
    J_{m}^{\mu}
    &= 
    J_{0}^{\mu}+J_{p}^{\mu}
    \, .
\end{split}
\end{align}
where $J_{0}^{\mu}$ sources the background EM field, and $J_{p}^{\mu}$ describes the plasma medium.

In the small coupling limit $g_{s \gamma} \phi_{0} \ll 1$ for the scalar, 
the coupling to the matter current in Eq.~\eqref{Eq: eom_elemag_scalar2} leads to
\begin{align}
\label{Eq: scalar_matter_coupling}
\begin{split}
    \frac{1}{1 + g_{s \gamma} \phi}  J_{m}^{\mu}
    &\approx 
    \left( 1 - g_{s \gamma} \phi \right) \left( J_{0}^{\mu} + J_{p}^{\mu} \right)
    \, .
\end{split}
\end{align}
The above is a unique result in the scalar case, and the different coupling to the matter current may also allow us to distinguish the scalar and ALP.
However, to focus only on the different couplings to the EM field in this work, we ignore the other matter coupling $g_{s\gamma}\phi_0 J^{\mu}_{0}$ even though it is the first order of perturbation.
We will revisit this unique coupling from the perspective of the scalar-tensor theory in Sec. 7.

The background electric and magnetic fields $\mathbf{E}_{0}$ and $\mathbf{B}_{0}$ are sourced by $J_{0}^{\mu} = (\rho_{0}, {\bf J}_{0})$.
For instance, the background matter fields can describe the magnetosphere of the neutron star or the intergalactic medium.
We note that $J_{0}^{\mu}$ is independent of the scalar or ALP field configuration.
We assume that the spatial extent of the scalar or ALP configuration is much smaller than the coherent length of the background matter fields.
Then, the background matter fields can be considered spatially constant.
The perturbed electric and magnetic fields $\mathbf{E}_{r}$ and $\mathbf{B}_{r}$ correspond to radiated EM fields, 
which is sourced by $J_{s}^{\mu}$ which depends on the background EM fields $\mathbf{E}_{0}$, $\mathbf{B}_{0}$, and the scalar field $\phi$.

Finally, we remark on the conservation law associated with the EM field.
Adopting the small coupling limit in the above discussion and dropping the coupling $g_{s\gamma}\phi_0 J^{\mu}_{0}$ for the scalar field, the Maxwell equation in both scalar and ALP cases yields the following conservation law approximately:
\begin{align}
    \partial_{\mu} \partial_{\nu}  F^{\mu\nu}
    &=
    0
    \approx
    \partial_{\nu}
    \left( J_{m}^{\nu} + J_{s}^{\nu} \right)
    \, .
\end{align}
Due to the coupling between the scalar/ALP field and EM field, the matter current $J^{\mu}_m$ is not solely conserved, and we have the total current conservation involving the scalar/ALP current $J^{\mu}_{s}$.
In the later consideration, we do not specify the analytic form of the matter current.
While we have assumed two configurations for the oscillating scalar/ALP fields in this subsection, it is in principle possible to keep the consistency in our consideration by adjusting the matter current $J^{\mu}_m$, so that we can satisfy the total EM conservation law.
In the next subsection, we consider the effect of the plasma medium $J^{\mu}_{p}$.

\subsection{Plasma medium effect}

Radiated EM fields can also be affected by the plasma medium $J^{\mu}_{p}$.
To investigate the plasma medium effect on the EM radiation, we consider the EM wave propagation through the background plasma with the modified dispersion relation
\begin{align}
    k^{2}= \omega^{2} - \frac{\omega_{p}^{2}\omega}{\omega + i \nu}
    \, .
\end{align}
$\omega$ and $\nu$ are the EM radiation frequency and collision frequency, respectively.
The plasma frequency $\omega_{p}$ is given by
\begin{align}
    \omega_{p} =\sqrt{\frac{4\pi n_{e} e^{2}}{m_{e}}}~,
\end{align}
where $m_{e}$ is the electron mass, $e$ is the electron's charge, and $n_{e}$ is the density of electrons.

In the collisionless limit, which is a good approximation to describe the hot plasma surrounding the compact stars and in the magnetosphere, $\omega \gg \nu$, the dispersion relation is given by 
\begin{align}
    k^{2}= \omega^{2} - \omega_{p}^{2}
    \, .
\end{align}
Since we assume the background EM fields are spatially constant in this work, 
$\omega_{p}$ is assumed to be a constant, though $\omega_{p}$ depends on the spatially varying free electron density.
In the following analysis, the plasma effect is incorporated into the perturbed equation through the modified dispersion relation.

\section{Scalar case}

In this section, we apply the perturbative approach established in the previous section to the pure scalar field, computing the EM radiation from the oscillating scalar field.
In this work, we consider three different background configurations of the scalar and EM fields.
After demonstrating a detailed calculation for the first setting, we present the results for the other two settings.
We particularly highlight the resonance effects, which can potentially enhance the observability of EM radiation, in the analytic expressions for the time-averaged EM radiation power.

From Eq.~\eqref{Eq: maxwell_scalar_pert}, the Maxwell equations for the background fields are given as
\begin{align}
\begin{split}
    \nabla \times \mathbf{B}_{0}\left(\mathbf{x}, t\right) 
    - \dot{\mathbf{E}}_{0} \left(\mathbf{x}, t\right)
    &=
    {\bf J}_{0} \left(\mathbf{x}, t\right)
    \, ,\\
    \nabla \times \mathbf{E}_{0} \left(\mathbf{x}, t\right)
    + \dot{\mathbf{B}}_{0} \left(\mathbf{x}, t\right)
    &= 0
    \, ,\\
    \nabla \cdot \mathbf{B}_{0} \left(\mathbf{x}, t\right)
    &= 0
    \, ,\\
    \nabla \cdot \mathbf{E}_{0} \left(\mathbf{x}, t\right)
    &= 
    \rho_{0} \left(\mathbf{x}, t\right)
    \, ,
\end{split}
\end{align}
and radiated EM fields obey the following equations
\begin{align}
\begin{split}
    \nabla \times \mathbf{B}_{r} \left(\mathbf{x}, t\right)
    - \dot{\mathbf{E}}_{r} \left(\mathbf{x}, t\right)
    &=
    {\bf J}_{p} \left(\mathbf{x}, t\right)
    + {\bf J}_{s} \left(\mathbf{x}, t\right)
    \, ,\\
    \nabla \times \mathbf{E}_{r} \left(\mathbf{x}, t\right)
    + \dot{\mathbf{B}}_{r} \left(\mathbf{x}, t\right)
    &= 
    0
    \, ,\\
    \nabla \cdot \mathbf{B}_{r} \left(\mathbf{x}, t\right)
    &= 0
    \, ,\\
    \nabla \cdot \mathbf{E}_{r} \left(\mathbf{x}, t\right)
    &= \rho_{p} \left(\mathbf{x}, t\right)
    + \rho_{s} \left(\mathbf{x}, t\right)
    \, .
\end{split}
\end{align}
The plasma effect on the propagation of EM waves induced from the terms $\rho_{p}$ and ${\bf J}_{p}$ can be written as the dispersion relation discussed in the previous subsection.
Thus, the EM radiation related to $\mathbf{E}_{r}$ and $\mathbf{B}_{r}$ is determined by the charge density and current of the scalar field $\rho_{s}$ and ${\bf J}_{s}$ coupled to the background EM fields:
\begin{align}
\label{Eq: charge_elemag_scalar1}
    \rho_{s} \left(\mathbf{x}, t\right)
    &= 
    -\frac{g_{s \gamma }}{1+g_{s \gamma }\phi \left(\mathbf{x}, t\right)} 
    \nabla \phi \left (\mathbf{x}, t\right) \cdot \mathbf{E}_{0} \left(\mathbf{x}, t\right)
    \, , \\ 
\label{Eq: 3current_elemag_scalar1}
    {\bf J}_{s} \left(\mathbf{x}, t\right)
    &= 
    \frac{g_{s \gamma }}{1+g_{s \gamma }\phi \left(\mathbf{x}, t\right)} 
    \left[
        \dot{\phi} \left(\mathbf{x}, t\right) \mathbf{E}_{0} \left(\mathbf{x}, t\right) 
        - \nabla \phi \left(\mathbf{x}, t\right) \times \mathbf{B}_{0} \left(\mathbf{x}, t\right)
    \right]
    \, .
\end{align}
We denote the frequency of the oscillating scalar field, its mass, and the frequency of the oscillating radius of the scalar-field condensate by $\omega_{s}$, $m_{s}$, and $\omega_{so}$.

Specifying the three functions $\{ \mathbf{E}_{0}, \mathbf{B}_{0}, \phi \}$, we can solve the differential equations for the EM radiation.
Although those background fields are essentially determined by the background Maxwell equation with the background matter current $J^{\mu}_{0}$,
we follow the previous studies~\cite{Amin:2021tnq,Sen:2021mhf} and test several background configurations as the benchmark.
In the following, we will consider three cases:
constant and alternating background magnetic field with Eqs.~\eqref{Eq: bkgscalar1}; alternating magnetic field with Eq.~\eqref{Eq: bkgscalar2}.
We apply these three settings to the scalar and ALP fields in this section and the next section.

\subsection{Constant magnetic field}
\label{Sec: scalar_const}

First, we consider a simple setup with a constant magnetic field.
We assume the background magnetic field is constant in the $z$ direction, and the background electric field vanishes,
\begin{align}
\label{Eq: scalar_const_EB}
\begin{split}
    \mathbf{E}_{0} \left (\mathbf{x}, t\right) 
    &= 0
    \, , \\
    \mathbf{B}_{0} \left(\mathbf{x}, t\right) 
    &= 
    B_{0}~\hat{\mathbf{E}}_{z}
    \, ,
\end{split}
\end{align}
Substituting Eqs.~\eqref{Eq: scalar_const_EB} and \eqref{Eq: bkgscalar1} into Eqs.~\eqref{Eq: charge_elemag_scalar1} and \eqref{Eq: 3current_elemag_scalar1}, we obtain the charge density and current,
\begin{align}
\begin{split}
    \rho_{s} \left (\mathbf{x}, t\right) 
    &= 0
    \, ,
\end{split}
\\
\label{current density-constant-pure scalar}
\begin{split}
    {\bf J}_{s}\left(\mathbf{x}, t\right) 
    &= 
    \frac{g_{s \gamma }}{1+g_{s \gamma }\phi \left ( \mathbf{x},t\right)} 
    \left[ 
        - \nabla \phi\left (\mathbf{x}, t\right) \times \mathbf{B}_{0} \left (\mathbf{x}, t\right) 
    \right]
    \\
    &\approx  
    \frac{g_{s \gamma} \phi_{0} B_{0}}{R} \cos\left(m_{s} t\right) 
    \mathrm{sech} \left( \frac{\left|\mathbf{x} \right|}{R}\right)
    \tanh \left( \frac{\left| \mathbf{x} \right|}{R}\right) 
     \hat{\mathbf{x}} \times \hat{\mathbf{E}}_{z}
    \, .
\end{split}
\end{align}
Here, we used $g_{s\gamma}\phi_{0} \ll 1$ in the second line of Eq.~\eqref{current density-constant-pure scalar}, and $\hat{\mathbf{x}}$ is the unit vector in the radial direction.
Using the above source, we solve the Maxwell equations and derive the radiated EM field $\mathbf{E}_{r}\left(\mathbf{x},t\right)$ and $\mathbf{B}_{r}\left(\mathbf{x},t\right)$.

In order to express the field equations in wave-equation form, we impose the Lorenz gauge condition
\begin{align}
    \partial_{\mu} A^{\mu}=0
    \, .
\end{align}
Under this condition, the Maxwell equation for the EM field $A^{\mu}\left(\mathbf{x},t\right)$ with a source $J^{\mu}\left(\mathbf{x},t\right)$ is written as
\begin{align}
    \Box A^{\mu} \left(\mathbf{x}, t\right) 
    = -J^{\mu} \left(\mathbf{x}, t\right)
    \, , 
\end{align}
where $A^{\mu} =\left(A^{0}, {\bf A}\right)$ and $J^{\mu} =\left(\rho, {\bf J}\right)$.
We can solve the above equation by the Green's function method, and the corresponding retarded Green's function is written as follows:
\begin{align}
\begin{split}
    G\left(\mathbf{x}, t; \mathbf{x}^{\prime}, t^{\prime}\right)
    &= -\frac{1}{2\pi}\int \frac{d \omega}{4\pi\left|\mathbf{x}-\mathbf{x}^{\prime}\right|} e^{-i\omega(t-t^{\prime})+ik_{s}\left|\mathbf{x}-\mathbf{x}^{\prime}\right|}\theta(k_{s}^{2}) 
    \\
    & \qquad
    -\frac{1}{2\pi}\int \frac{d \omega}{4\pi\left|\mathbf{x}-\mathbf{x}^{\prime}\right|} e^{-i\omega(t-t^{\prime})-\sqrt{|k_{s}^{2}|}\left|\mathbf{x}-\mathbf{x}^{\prime}\right|}\theta(-k_{s}^{2})
    \, ,
\end{split}
\end{align}
$k_s$ includes the plasma effect, $k_{s}=\sqrt{m_{s}^{2}-\omega_{p}^{2}}$, 
and $\theta(k_{s}^{2})$ is the Heaviside step function.
If the frequency of EM radiation is smaller than that of the plasma ($\omega \sim m_{s} < \omega_{p}$), the radiation is exponentially damped. 
Furthermore, we assume the plasma frequency is smaller than the scalar field mass.

The vector potential ${\bf A}\left(\mathbf{x}, t\right)$ is expressed in terms of the Green's function as
\begin{align}
\label{Eq: EMfield}
\begin{split}
    &{\bf A}\left(\mathbf{x}, t\right) 
    \\
    &= 
    \int d^{3}\mathbf{x}^{\prime} d t^{\prime} 
    G\left(\mathbf{x}, t; \mathbf{x}^{\prime}, t^{\prime}\right) 
    {\bf J}_{s}\left(\mathbf{x}^{\prime}, t^{\prime}\right) 
    \\
    &=
    \frac{g_{s\gamma}\phi_{0}B_{0}}{4\pi R}
    \left[
        \int_{0}^{\infty}|\mathbf{x}^{\prime}|^{2}d |\mathbf{x}^{\prime}| \int_{0}^{\pi} 
        \sin \theta^{\prime}d\theta^{\prime} 
        \int_{0}^{2\pi}d\varphi^{\prime} 
        \frac{1}{\left|\mathbf{x}-\mathbf{x}^{\prime}\right|} 
    \right.
    \\
    & \qquad \qquad \qquad \quad 
    \left.
        \cos \left( m_{s} t- k_{s}\left|\mathbf{x}-\mathbf{x}^{\prime}\right| \right) 
        {\rm sech} \left( \frac{|\mathbf{x}^{\prime}|}{R} \right)
        \tanh \left( \frac{|\mathbf{x}^{\prime}|}{R} \right)
    \right]
    \hat{\mathbf{x}} \times \hat{\mathbf{E}}_{z} 
    \\
    &\approx 
    \frac{g_{s\gamma}\phi_{0}B_{0}}{2 |\mathbf{x}| R} 
    \left\{
        \int_{0}^{\infty}|\mathbf{x}^{\prime}|^{2}d |
        \mathbf{x}^{\prime}|\int_{-1}^{1}d(\cos\theta^{\prime}) 
    \right.
    \\
    & \qquad \qquad \qquad \quad 
    \left.
        \cos \left[m_{s} t - k_{s} \left( |\mathbf{x}| - |
        \mathbf{x}^{\prime}|\cos\theta^{\prime} \right) \right] 
    \right.
    \\
    & \qquad \qquad \qquad \qquad \qquad \qquad \qquad 
    \left.    
        {\rm sech} \left( \frac{|\mathbf{x}^{\prime}|}{R} \right)
        \tanh \left( \frac{|\mathbf{x}^{\prime}|}{R} \right)
    \right\}
    \hat{\mathbf{x}} \times \hat{\mathbf{E}}_{z}
    \, .
\end{split}
\end{align}
$\theta^{\prime}$ is the angle between $\mathbf{x}$ and $\mathbf{x}^{\prime}$, and we used an approximation
\begin{align}
    |\mathbf{x}-\mathbf{x}^{\prime}| 
    \approx 
    |\mathbf{x}|-\frac{\mathbf{x} \cdot \mathbf{x}^{\prime}}{|\mathbf{x}|}= |\mathbf{x}| - |\mathbf{x}^{\prime}|\cos\theta^{\prime}
\end{align}
in the third equality of Eq.~\eqref{Eq: EMfield}.
In the above calculation, we consider the coordinate system where the $\hat{\mathbf{E}}^{\prime}_{z}$ coincides with the direction of an observer ($\hat{\mathbf{x}}$).

The radiated electric and magnetic fields are written by the corresponding EM field $A^{\mu}$ as 
\begin{align}
\begin{split}
    \mathbf{E}_{r}\left(\mathbf{x}, t\right)
    &=
    - \nabla A_{0}\left(\mathbf{x}, t\right)- \partial_{t} {\bf A}\left(\mathbf{x}, t\right)
    \, , \\ 
    \mathbf{B}_{r}\left(\mathbf{x}, t\right)
    &= 
    \nabla \times {\bf A}\left(\mathbf{x}, t\right)
    \, , 
\end{split}
\end{align}
and the Poynting flux for the radiated electric and magnetic field is defined by 
\begin{align}    
    {\bf S} \left(\mathbf{x}, t\right)
    =
    \mathbf{E}_{r}\left(\mathbf{x}, t\right) \times \mathbf{B}_{r}\left(\mathbf{x}, t\right)
    \, .
\end{align}
Since the radiated power per unit solid angle is expressed as
\begin{align}
    \label{Eq: power_per_unit}
    \frac{d P}{d\Omega} =|\mathbf{x}|^{2} {\bf S} \cdot \hat{\mathbf{x}}
    \, ,
\end{align}
we obtain the time-averaged radiation power as 
\begin{align}
\label{Eq: pta_scalar}
    \mathcal{P}
    =
    |\mathbf{x}|^{2} \int \bar{\bf S} \cdot \hat{\mathbf{x}} d\Omega
    =4\pi|\mathbf{x}|^{2} |\bar{\bf S}|
    \,,
\end{align}
where $\bar{\bf S}=\frac{1}{T}\int_{0}^{T} {\bf S} dt$ and $T=\frac{2\pi}{m_{s}}$.
We will show the details in Appendix~A
In the following two subsections, this result will be applied to the other two cases involving alternating magnetic fields.

Regarding the differential power per solid angle in Eq.~\eqref{Eq: power_per_unit}, since the scalar field configurations considered here are spherically symmetric, the angular dependence of the radiation is entirely determined by the background magnetic field, which defines a preferred direction (the z-axis as in Eq.~\eqref{Eq: scalar_const_EB} in our analysis).
As a result, the radiation pattern is axially symmetric around the magnetic field direction.
The emission is suppressed along the z-axis and enhanced in the transverse plane, resulting in a dipole-like angular structure.
For both cases of the oscillating field configuration in Eq.~\eqref{Eq: bkgscalar1} and time-dependent radius in Eq.~\eqref{Eq: bkgscalar2}, this qualitative feature remains unchanged, while the frequency spectrum differs depending on the temporal behavior of the scalar configuration.
In the present work, we focus on the total emitted power and leave a detailed quantitative analysis of the angular distribution to future investigation.

According to the above definitions and relations, the EM radiation emitted from the scalar field in the presence of a constant external magnetic field is obtained as
\begin{align}
\label{radiation-power-constant}
\begin{split}
    \mathcal{P}
    &= 
    \frac{1}{128 k_{s}} \pi g_{s \gamma}^{2} \phi_{0}^{2} B_{0}^{2} R^{2} m_{s} \left[4i Q^0_{-}(k_s R)+k_s R Q^{1}_+(k_s R)\right]^2
    \\
    &=
    \frac{ \pi g_{s \gamma}^{2} \phi_{0}^{2} B_{0}^{2}}{128 m_{s}^2\sqrt{1-\xi_{s}^2}}\ R_{s*}^{2}  
    \\
    & \qquad \times
    \left[
        4i Q^0_{-}(\sqrt{1-\xi_{s}^2} R_{s*})
        +\sqrt{1-\xi_{s}^2} R_{s*} Q^{1}_+(\sqrt{1-\xi_{s}^2}R_{s*})
    \right]^2
    \, ,
\end{split}
\end{align}
where $R_{s*}$ is the dimensionless size variable $R_{s*}=m_s R$.
$\xi_{s}$ is newly introduced parameter $\xi_{s}=\omega_p /m_{s}$, which is helpful discuss the plasma medium effect.
And, the function $Q^{n}_{\pm}$ is defined as
\begin{align}
\begin{split}
    Q^{n}_{\pm}(x)
    &:=
    \psi^{(n)} \left(\frac{1}{4}-i\frac{ x}{4}\right)
    -\psi^{(n)} \left(\frac{3}{4}-i\frac{ x}{4}\right) 
    \\
    & \qquad
    \pm \psi^{(n)} \left(\frac{1}{4}+i\frac{ x }{4}\right)
    \mp \psi^{(n)} \left(\frac{3}{4}+i\frac{ x}{4}\right)
\end{split}
\end{align}
with the Polygamma function by $\psi^{n}(x)$, which arises from the integral in Eq.~\eqref{Eq: EMfield}. 
Note that for a real variable $x$, $Q^0_\pm(x)$ and $Q^1_\pm(x)$ are purely imaginary and real functions, respectively.
Therefore, $\mathcal{P}$ is a real function of $R_*$, and we will discuss the behavior of radiated power in Sec.~5.

It is clear from Eq.~\eqref{radiation-power-constant} that the radiated power depends on the size of $R_{s*}$.
When the plasma medium effect is negligible ($\omega_{p}=0$), 
the radiation exhibits peaks for $R_{s*} \sim 0.84$, whereas the radiation is suppressed for $R_{s*} \gg 0.84$.
When $\omega_{p}$ is non-negligible, the critical value becomes 
\begin{align}
    R_{s*} \sim \frac{0.84}{\sqrt{1 - \xi_{s}^{2}}}
    \, .
\end{align}
Thus, the resonant effect occurs when plasma frequency is close to the scalar mass scale ($\xi_s = \omega_p /m_{s} \sim 1$).
This resonant effect can enhance the radiated power when the size of the scalar field $R$ is much larger than the inverse of the scalar-field mass scale $m^{-1}_{s}$
($R_{s*} = R/m^{-1}_{s}\gg1$).
For instance, we can expect the enhanced EM radiation from the compact scalar-field condensate for the massive scalar field.

However, the resonant effect does not always enhance the radiated power.
For $R_{s*} \ll 1 / \sqrt{1 - \xi_{s}^{2}}$, the time-averaged radiation power $\mathcal{P}$ yields
 \begin{align}
 \label{approx}
      \mathcal{P}
     \approx 
      \frac{\pi g_{s \gamma}^{2} \phi_{0}^{2} B_{0}^{2} }{ 32 m_{s}^{2}} \sqrt{1 - \xi_{s}^{2}}  R_{s*}^{4}   
      \left[\psi^{(1)}(1/4)-\psi^{(1)}(3/4) \right]^2
      \, .
 \end{align}
It shows that radiation generated under the condition of $\omega_{p} = 0$ is stronger than that generated under the resonance condition ($m_{s} \sim \omega_{p}$).
In the following, we will explore a similar resonant effect that can occur in an alternating magnetic field background.

For comparison, we consider the synchrotron radiation power of a relativistic charged particle in a constant magnetic field $B_0$.
In this case, the synchrotron radiation power $\mathcal{P}_{\mathrm{syn}}$ is given by
\begin{align}
    \mathcal{P}_{\mathrm{syn}} 
    = 
    \frac{2}{3} \frac{e^4}{m^2} B_0^2 \gamma^2
    \, ,
\end{align}
where $e$ and $m$ denote the electric charge and mass of the radiating particle, respectively, and $\gamma$ is the Lorentz factor. 
Both synchrotron and scalar-induced radiation exhibit a quadratic dependence on the magnetic field strength $B_0$. 
However, synchrotron radiation is enhanced by relativistic particle motion, whereas scalar-induced radiation is governed by the scalar–photon coupling strength and the properties of the coherent scalar configuration.

\subsection{Alternating magnetic field}
\label{Sec: scalar_alternating}

Next, we consider the alternating magnetic field as the background, which may appear around the spinning neutron stars~\cite{Pons:2007vf, PhysRevLett.112.171101}.
The background electric and magnetic fields are assumed to be
\begin{align}
\label{Eq: scalar_alternating_EB}
\begin{split}
    \mathbf{E}_{0} \left (\mathbf{x}, t\right) 
    &= 0
    \, , \\
    \mathbf{B}_{0} \left(\mathbf{x}, t\right) 
    &= 
    B_{0} \cos\left(\Omega t\right) \hat{\mathbf{E}}_{z}
    \, ,
\end{split}
\end{align}
where $\Omega$ is the frequency of the magnetic field, and we ignore the initial phase shift of the frequency. 
Substituting Eqs.~\eqref{Eq: scalar_alternating_EB} and \eqref{Eq: bkgscalar1} into Eqs.~\eqref{Eq: charge_elemag_scalar1} and \eqref{Eq: 3current_elemag_scalar1}, we can express the charge density and current as
\begin{align}
\begin{split}
    \rho_{s}\left (\mathbf{x}, t\right) 
    &= 0
    \, .
\end{split}
    \\
\begin{split}
    {\bf J}_{s}\left(\mathbf{x}, t\right) 
    &\approx 
    \frac{g_{s \gamma} \phi_{0} B_{0}}{R} \cos\left(m_{s} t\right) 
    {\rm sech}\left( \frac{\left|\mathbf{x} \right|}{R} \right)
    \tanh \left(\frac{\left|\mathbf{x} \right|}{R}\right) 
    \cos \left(\Omega t \right ) \hat{\mathbf{x}} \times \hat{\mathbf{E}}_{z}
    \, , 
\end{split}
\end{align}
As in the previous subsection, we compute the time-averaged radiated power
and obtain the following expression in both $\omega_{s} > \Omega$ and $\Omega > \omega_{s}$ cases,
\begin{align}
    \label{radiation-power-alternating}
\begin{split}
    \mathcal{P} 
    &=
    \dfrac{\pi g_{s \gamma}^{2} \phi_{0}^{2} B_{0}^{2}}{256 m_{s}^{2}} R_{s*}^{2}
    \\
    & \quad \times
    \left\{ 
    \dfrac{1 + \zeta_{s}}{f_{s+}} \left[4iQ^0_{-}(f_{s+} R_{s*})
    + f_{s+}R_{s*} Q^{1}_+(f_{s+} R_{s*}) \right]^2 \theta(f_{s+}^2)  
    \right. \\
    & \quad \qquad \left.
    + \dfrac{1 - \zeta_{s}}{f_{s-}} \left[4iQ^0_{-}(f_{s-} R_{s*})+ f_{s-} R_{s*} Q^{1}_+(f_{s-} R_{s*}) \right]^2 \theta(f_{s-}^2)
    \right\}
    \, , 
\end{split}
\end{align}
where $f_{s\pm} = \sqrt{(1 \pm \zeta_{s})^{2} - \xi_{s}^{2}}$ and $\xi_{s}=\omega_p /m_{s}$.
$\zeta_{s}$ is a newly defined parameter $\zeta_{s} = \Omega/m_{s}$, which is analogous to $\xi_s$ and helpful discuss the alternating-magnetic field effect.

We find that Eq.~\eqref{radiation-power-alternating} reduces Eq.~\eqref{radiation-power-constant} when the magnetic field is constant in time $\Omega = 0$ ($\zeta_{s}=0$, and that the radiation is suppressed when $f_{ \pm}^{2} < 0$. 
The radiation has peaks for 
\begin{align}
    R_{s*} \sim \frac{0.84}{\sqrt{(1 \pm \zeta_{s})^{2} - \xi_{s}^{2}}}
    \, ,
\end{align}
and thus, the resonance effect depends on both the $\omega_{p}$ and $\Omega$.
The resonance effect appears when the plasma medium effect is negligible ($\omega_{p} = 0$ and $\xi_{s}=0$) and the scalar-field mass is close to the frequency of the alternating magnetic field $m_{s} \sim \Omega$ ($\zeta_{s} = \Omega/m_{s} \approx 1$), and, this resonant effect can also enhance the radiated power.

For $R_{s*} \ll 1 /\sqrt{(1 \pm \zeta_{s})^{2} - \xi_{s}^{2}}$, Eq.~\eqref{radiation-power-alternating} is reduced to
\begin{align}
\label{approx_alt}
\begin{split}
    \mathcal{P}
    &\approx 
    \frac{\pi g_{s \gamma}^{2} \phi_{0}^{2} B_{0}^{2}}{ 64 m_{s}^{2}} R_{s*}^{4} 
    \left[\psi^{(1)}(1/4)-\psi^{(1)}(3/4) \right]^2
    \\
    & \qquad \times
    \left[ 
        (1 + \zeta_{s})f_{s+} \theta(f_{s+}^{2})
        + (1 - \zeta_{s}) f_{s-} \theta(f_{s-}^{2})
    \right] 
    \, .
\end{split}
\end{align}
The above expression shows that
the scalar-field oscillation in resonance effect ($m_{s} \sim \Omega$, $\zeta_{s} = \Omega/m_{s} \approx 1$ and $\xi_{s}=0$) can emit EM radiation more efficiently than other resonance effects that occur when $m_{s} \sim \omega_{p}$ ($\xi_{s}=\omega_p /m_{s} \approx 1$) and $\Omega =0$ ($\zeta_{s} = 0$).

\subsection{Radial oscillation of scalar condensate}
\label{Sec: scalar_oscillation}

Finally, we consider the case in which the radius of the scalar field oscillates. 
As in Eq.~\eqref{Eq: scalar_alternating_EB}, we consider the alternating background magnetic field and assume the background electric field vanishes.
Moreover, we ignore the plasma effect for simplicity.
Substituting Eqs.~\eqref{Eq: scalar_alternating_EB} and \eqref{Eq: bkgscalar2} into Eqs.~\eqref{Eq: charge_elemag_scalar1} and \eqref{Eq: 3current_elemag_scalar1}, the charge density and current are expressed as
\begin{align}
\begin{split}
    \rho_{s}\left (\mathbf{x}, t\right) 
    &= 0
    \, ,
\end{split}
    \\
\begin{split}
\label{current density-oscillation-pure scalar}
    {\bf J}_{s}\left(\mathbf{x}, t\right) 
    &\approx  
    \frac{ g_{s \gamma } \phi_{0} B_{0} \cos\left[\Omega t \right ]}
    {R \left(1 + \delta_R \cos [\omega_{so} t] \right)}  
    {\rm sech} 
    \left[
        \frac{ \left|\mathbf{x} \right|}
        {R \left(1 + \delta_R \cos [\omega_{so} t] \right)}
    \right]
    \\
    & \qquad \quad \times
    \tanh
    \left[
        \frac{ \left|\mathbf{x} \right|}
        {R \left(1 + \delta_R \cos [\omega_{so} t] \right)}
    \right] 
    \hat{\mathbf{x}} \times \hat{\mathbf{E}}_{z}
    \, . 
\end{split}
\end{align}

We note that in Eq.~\eqref{current density-oscillation-pure scalar},
$\delta_R = 0$ leads to the current density in the case of the constant magnetic field, where the frequency of the alternating magnetic field mimics the role of the frequency of the oscillating scalar field in Eq.~\eqref{current density-constant-pure scalar}.
Consequently, the generated EM radiation has the same characteristics as in the case of a constant magnetic field, where $\omega_{p}$ is negligible.
Because in the current setup, we cannot obtain the analytic form of the time-averaged radiated power for $\delta_R  \ll 1$, we will show the numerical results in Sec.~5.

\section{ALP case}

In this section, we apply the three background settings for the scalar field in the previous section to the ALP field and analyze the radiation power for comparison. 
After discussing the calculation method for the ALP field, which differs from that for the scalar field due to its different coupling to the EM field, we present the results in the three settings.
We also discuss the resonance effects in the analytic expressions for the ALP field.

From Eq.~\eqref{Eq: maxwell_axion_pert}, the Maxwell equations for the background field are given as
\begin{align}
\begin{split}
    \nabla \times \mathbf{B}_{0}\left(\mathbf{x}, t\right) 
    - \dot{\mathbf{E}}_{0} \left(\mathbf{x}, t\right)
    &=
    {\bf J}_{0} \left(\mathbf{x}, t\right)
    \, ,\\
    \nabla \times \mathbf{E}_{0} \left(\mathbf{x}, t\right)
    + \dot{\mathbf{B}}_{0} \left(\mathbf{x}, t\right)
    &= 0
    \, ,\\
    \nabla \cdot \mathbf{B}_{0} \left(\mathbf{x}, t\right)
    &= 0
    \, ,\\
    \nabla \cdot \mathbf{E}_{0} \left(\mathbf{x}, t\right)
    &= 
    \rho_{0} \left(\mathbf{x}, t\right)
    \, ,
\end{split}
\end{align}
and radiated EM fields are given as
\begin{align}
\begin{split}
    \nabla \times \mathbf{B}_{r} \left(\mathbf{x}, t\right)
    - \dot{\mathbf{E}}_{r} \left(\mathbf{x}, t\right)
    &=
    {\bf J}_{p} \left(\mathbf{x}, t\right)
    + {\bf J}_{s} \left(\mathbf{x}, t\right)
    \, ,\\
    \nabla \times \mathbf{E}_{r} \left(\mathbf{x}, t\right)
    + \dot{\mathbf{B}}_{r} \left(\mathbf{x}, t\right)
    &= 
    0
    \, ,\\
    \nabla \cdot \mathbf{B}_{r} \left(\mathbf{x}, t\right)
    &= 0
    \, ,\\
    \nabla \cdot \mathbf{E}_{r} \left(\mathbf{x}, t\right)
    &= \rho_{p} \left(\mathbf{x}, t\right)
    + \rho_{s} \left(\mathbf{x}, t\right)
    \, .
\end{split}
\end{align}
As in the case of the scalar field, we drop the terms $\rho_{p}$ and ${\bf J}_{p}$ in the above equations and evaluate the plasma effect as the modified dispersion relation.
The EM radiation depends on the charge density and current of the ALP field coupled to the background EM fields,
\begin{align}
\label{Eq: charge_elemag_axion1}
    \rho_{s} \left(\mathbf{x}, t\right)
    &= 
    -g_{a \gamma } \nabla \phi\left (\mathbf{x}, t\right) \cdot \mathbf{B}_{0} \left(\mathbf{x}, t\right)
    \, , \\ 
\label{Eq: 3current_elemag_axion1}
    {\bf J}_{s} \left(\mathbf{x}, t\right)
    &= 
    g_{a \gamma } 
    \left[
        \dot{\phi} \left(\mathbf{x}, t\right) \mathbf{B}_{0} \left(\mathbf{x}, t\right) 
        + \nabla \phi \left(\mathbf{x}, t\right) \times \mathbf{E}_{0} \left(\mathbf{x}, t\right)
    \right]
    \, .
\end{align}

To demonstrate the qualitative comparison with the scalar-field case, we consider the EM radiation generated by the ALP field under the same three settings as in the previous section.
We denote the frequency of the oscillating ALP field, its mass, and the frequency of the oscillating radius of the ALP-field condensate by $\omega_{a}$, $m_{a}$, and $\omega_{ao}$ for the ALP field.

\subsection{Constant magnetic field}
\label{Sec: axion_const}

First, we apply the settings in Sec~\ref{Sec: scalar_const} to the ALP field.
Substituting Eqs.~\eqref{Eq: scalar_const_EB} and \eqref{Eq: bkgscalar1} into Eqs.~\eqref{Eq: charge_elemag_axion1} and \eqref{Eq: 3current_elemag_axion1}, 
we obtain
\begin{align}
\begin{split}
    \rho_{s}\left (\mathbf{x}, t\right) 
    &= 
    -g_{a\gamma} \nabla \phi\left (\mathbf{x}, t\right) \cdot \mathbf{B}_{0}\left(\mathbf{x}, t\right)  
    \\
    &\approx 
    \frac{g_{a \gamma }\phi_{0} B_{0}}{R} \cos(m_{a} t) 
    \mathrm{sech} \left( \frac{\left|\mathbf{x} \right|}{R} \right)
    \tanh \left( \frac{\left|\mathbf{x} \right|}{R} \right)
    \hat{\mathbf{x}} \cdot \hat{\mathbf{e}}_{z}
    \, ,
\end{split}
    \\
\begin{split}
    {\bf J}_{s}\left(\mathbf{x}, t\right) 
    &= 
    g_{a \gamma } \dot{\phi}\left(\mathbf{x}, t\right)  \mathbf{B}_{0}\left(\mathbf{x}, t\right) 
    \\
    &\approx  
    - g_{a \gamma } m_{a} \phi_{0} B_{0} \sin\left(m_{a} t\right) 
    \mathrm{sech} \left( \frac{\left|\mathbf{x} \right|}{R} \right)
    \hat{\mathbf{E}}_{z}
    \, , 
\end{split}
\end{align}
It is worth mentioning that the charge density does not vanish in the ALP case despite the same background field configurations.
Applying the Green's function method, we obtain the radiated EM field $A^{\mu}$.
However, the time-averaged radiation power in the case of ALP is different from that of the scalar due to the different coupling to the EM field,
\begin{align}
\label{Eq: pta_axion}
    \mathcal{P}
    =
    |\mathbf{x}|^{2} \int \bar{\bf S} \cdot \hat{\mathbf{x}} d\Omega
    =\frac{8}{3}\pi|\mathbf{x}|^{2} |\bar{\bf S}|
    \, .
\end{align}
where $\bar{\bf S}=\frac{1}{T}\int_{0}^{T} {\bf S} dt$ and $T=\frac{2\pi}{m_{a}}$.
This result will again apply to the other two cases for alternating magnetic fields in the following two subsections.

Based on the above consideration, the time-averaged radiated power in the presence of a constant external magnetic field and plasma is given by
\begin{align}
    \mathcal{P} 
    = 
    \frac{4 \pi^{5} g_{a \gamma}^{2} \phi_{0}^{2} B_{0}^{2}}{3 m_{a}^{2} \sqrt{1 - \xi_{a}^{2}}} R_{a*} ^{4}
    \mathrm{csch}\left(\pi \sqrt{1 - \xi_{a}^{2}} R_{a*} \right)^{4} 
    \sinh\left(\frac{\pi \sqrt{1 - \xi_{a}^{2}} R_{a*}}{2}\right)^{6}~,
\end{align}
where $\xi_{a}= \omega_{p}/m_{a}$, $R_{a*} = m_{a} R$.
We find that the above analytic form of the radiation power for the ALP field is quite different from that for the scalar field as in Eq.~\eqref{radiation-power-constant}, even though the settings for the background fields are the same.
We find that different couplings to the EM field indeed lead to different currents, resulting in distinct analytic expressions of the radiation power.
We note that the above result is consistent with the existing works~\cite{Amin:2021tnq, Sen:2021mhf}.
When $\omega_p$ is non-negligible, the radiated power peaks for the case that 
\begin{align}
    R_{a*}  \sim \frac{1.38}{\sqrt{1 -\xi_{a}^{2}}}
    \, ,
\end{align}
and when $R_{a*} \gg 1.38/\sqrt{1 -\xi_{a}^{2}} $, the radiated power is exponentially suppressed.
Up to the numerical factor, the peak structure of the radiation power in the ALP-field case is the same as that in the scalar-field case.

\subsection{Alternating magnetic field}
\label{Sec: axion_alternating}

Next, we apply the settings in Sec~\ref{Sec: scalar_alternating} to the ALP field.
Substituting Eqs.~\eqref{Eq: scalar_alternating_EB} and \eqref{Eq: bkgscalar1} into Eqs.~\eqref{Eq: charge_elemag_axion1} and \eqref{Eq: 3current_elemag_axion1}, we obtain the charge density and current,
\begin{align}
\begin{split}
    \rho_{s}\left (\mathbf{x}, t\right) 
    &\approx 
    \frac{g_{a \gamma }\phi_{0}}{R} \cos\left(m_{a} t\right) 
    \mathrm{sech}\left( \frac{\left|\mathbf{x} \right|}{R}\right)
    \tanh \left( \frac{\left|\mathbf{x} \right|}{R} \right) B_{0} 
    \cos\left(\Omega t \right) \hat{\mathbf{x}} \cdot \hat{\mathbf{E}}_{z}
    \, .
\end{split}
    \\
\begin{split}
    {\bf J}_{s}\left(\mathbf{x}, t\right) 
    &\approx  
    - g_{a \gamma } m_{a} \phi_{0} \sin\left(m_{a} t\right) 
    \mathrm{sech} \left[\frac{\left|\mathbf{x} \right|}{R}\right] B_{0} \cos\left(\Omega t\right) \hat{\mathbf{E}}_{z}
    \, , 
\end{split}
\end{align}
The corresponding time-averaged radiated power is given by
\begin{align}
\label{alternating B radiated power}
\begin{split}
    \mathcal{P}
    &=
    \frac{2 \pi^{5}  g_{a \gamma}^{2} \phi_{0}^{2} B_{0}^{2} }{3 m_{a}^{2}} R_{a*} ^{4}  
    \left[
        \frac{1 + \zeta}{f_{a+}}{\rm csch}\left(\pi f_{a+} R_{a*} \right)^{4} 
        \sinh\left(\frac{\pi f_{a+} R_{a*} }{2}\right)^{6} \theta(f_{a+}^{2}) 
    \right. \\
    & \qquad 
    \left.
        + \frac{1 - \zeta}{f_{a-}}{\rm csch}\left(\pi f_{a-} R_{a*} \right)^{4}
        \sinh\left(\frac{\pi f_{a-} R}{2}\right)^{6} \theta(f_{a-}^{2})  
    \right]
    \, ,
\end{split}
\end{align}
where $f_{a\pm} = \sqrt{(1 \pm \zeta_{a})^{2} - \xi_{a}^{2}}$, $\zeta_{a} = \Omega /m_{a}$ and the radiation be exponentially suppressed when $f_{a \pm} $ is not real. 
Eq.~\eqref{alternating B radiated power} shows that the radiation peaks for two different values of the field radius: 
\begin{align}
    R_{a*}  
    \sim 
    \frac{1.38}{\sqrt{(1 \pm \zeta_{a})^{2} - \xi_{a}^{2}}}
    \, .
\end{align} 
As in the constant-magnetic-field case, the expression of the radiation power is different from that for the scalar field, but the peak structure is the same as that in the scalar-field case up to the numerical factor.

\subsection{Radial oscillation of ALP condensate}
\label{Sec: axion_oscillation}

Finally, we apply the settings in Sec~\ref{Sec: scalar_oscillation} to the ALP field.
Substituting Eqs.~\eqref{Eq: scalar_alternating_EB} and \eqref{Eq: bkgscalar2} into Eqs.~\eqref{Eq: charge_elemag_axion1} and \eqref{Eq: 3current_elemag_axion1}, we obtain the charge density and current, 
\begin{align}
\label{current density-oscillation-axion}
\begin{split}
    &\rho_{s}\left (\mathbf{x}, t\right) 
    \\
    &\approx 
    \frac{ g_{a \gamma }\phi_{0}}{R \left(1 + \delta_R \cos [\omega_{ao} t] \right)} 
    \\
    & \qquad \times
    \mathrm{sech} \left[\frac{ \left|\mathbf{x} \right|}{R \left(1 + \delta_R \cos [\omega_{ao} t] \right)}\right] 
    \tanh \left[\frac{ \left|\mathbf{x} \right|}{R \left(1 + \delta_R \cos [\omega_{ao} t] \right)} \right] B_{0}  \hat{\mathbf{x}} \cdot \hat{\mathbf{E}}_{z}
    \, .
\end{split}
    \\
\begin{split}
    &{\bf J}_{s}\left(\mathbf{x}, t\right) 
    \\
    &\approx  
    -\frac{ \delta_{R}~\omega_{ao} \left|\mathbf{x} \right| g_{s \gamma } \phi_{0} \sin\left[\omega_{ao} t\right] \cos\left[\Omega t\right]}
    {R \left(1 + \delta_R \cos [\omega_{ao} t] \right)^{2}} 
    \\
    & \qquad \times
    {\rm sech}\left[\frac{ \left|\mathbf{x} \right|}{R \left(1 + \delta_R \cos [\omega_{ao} t] \right)}\right]
    \tanh\left[\frac{ \left|\mathbf{x} \right|}{R \left(1 + \delta_R \cos [\omega_{ao} t] \right)}\right] B_{0}\hat{\mathbf{E}}_{z}
    \, , 
\end{split}
\end{align}
We note that the current vanishes when $\delta_R = 0$, ${\bf J}_{s}\left(\mathbf{x}, t\right) \approx 0$.
Thus, we require $\delta_R \neq 0$ for the ALP field to radiate in the above background field setups.
We will discuss the numerical results for nonzero $\delta_R$ in Sec.~5.

\section{Comparison: scalar vs. ALP}

In this section, we analyze the qualitative behavior of the EM radiation powers generated by the scalar/ALP fields, utilizing the results from the previous two sections.
Through numerical calculations, we discuss the differences in the resonance effects of the EM radiation from scalar and ALP fields.
We also consider the detectability of EM signals generated by scalar/ALP fields with reasonable parameter settings.

\subsection{Qualitative behaviors of radiation power}

Based on the theoretical analysis presented in the previous two sections, we present numerical results for the EM radiation power from the scalar and ALP fields in three different cases and compare them. 
We plot these three cases for the scalar/ALP fields and discuss the resonant effects.

\begin{figure}[htbp]
    \begin{minipage}[b]{0.48\columnwidth}
    \centering
    \includegraphics[width=\columnwidth]{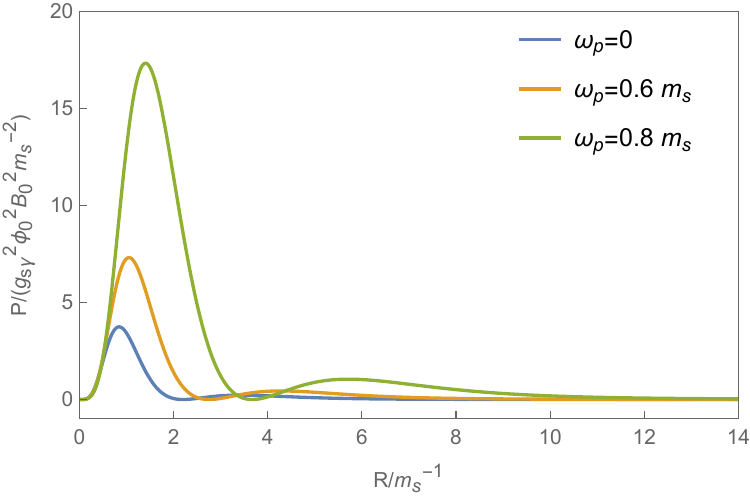}
    \end{minipage}
    \begin{minipage}[b]{0.48\columnwidth}
    \centering
    \includegraphics[width=\columnwidth]{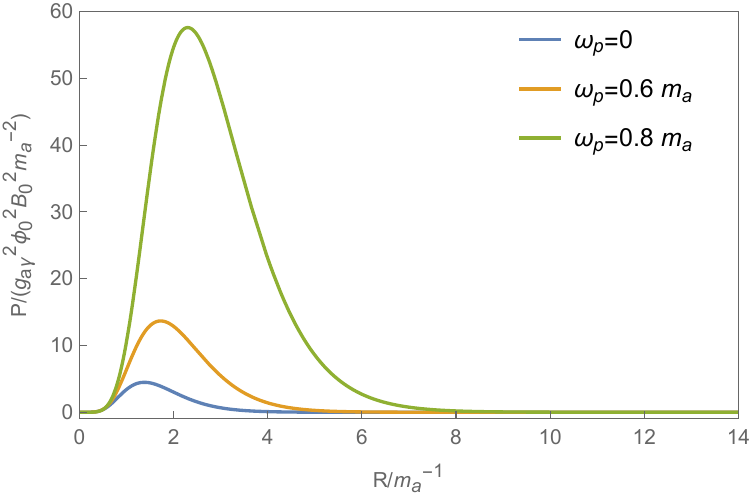}
    \end{minipage}
    \caption{
    The radiated power as a function of $R_*=R/m^{-1}$ in the case of the constant magnetic field for the scalar (Left panel) and ALP (Right panel).
    } 
    \label{Fig: const_mag}
\end{figure}

First, we plot the radiated power in the constant magnetic field in Fig.~\ref{Fig: const_mag}.
we consider three different values of the plasma frequency $\omega_p$, $\omega_p/m = 0, 0.6, 0.8$, for both the scalar and ALP cases.
As we mentioned in the Sec.~\ref{Sec: scalar_const} and \ref{Sec: axion_const}, peaks in the radiation power for the scalar and ALP are characterized by $R_* \sim C_{i}/\sqrt{1 - \xi^{2}}$, where $R_*$ denotes the product of the mass of scalar/ALP $m$ with the size of the field configuration $R$.
$C_{i}$ is a constant, and we denote it by $C_{s}$ and $C_{a}$ for the scalar and ALP fields, respectively.

\begin{figure}[htbp]
    \begin{minipage}[b]{0.48\columnwidth}
        \centering
        \includegraphics[width=\columnwidth]{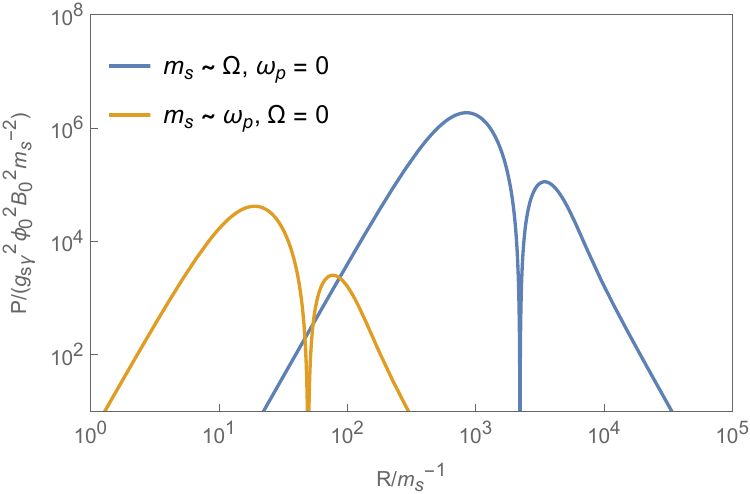}
    \end{minipage}
    \begin{minipage}[b]{0.48\columnwidth}
    \centering
    \includegraphics[width=\columnwidth]{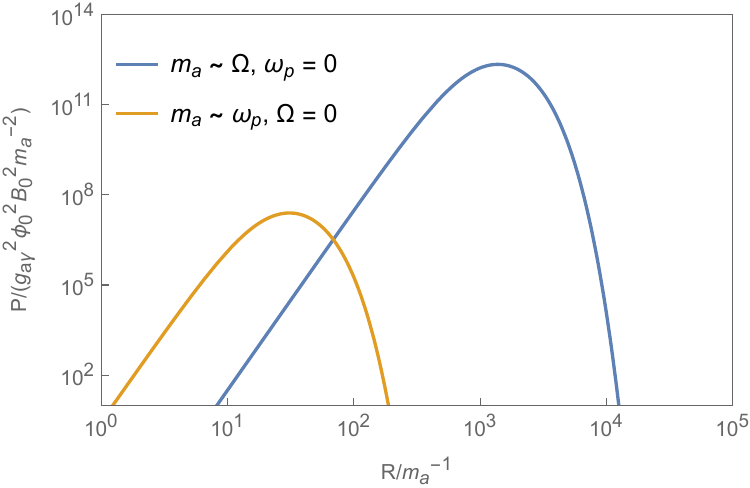}
    \end{minipage}
    \caption{
    The radiated power as a function of $R_*=R/m^{-1}$ under the resonance effect for the scalar (Left panel) and ALP (Right panel).
    }
    \label{Fig: alt_mag_R}
\end{figure}

Fig.~\ref{Fig: const_mag} shows that values of $R_*$ at the radiation peak for scalar and ALP are not identical as in the theoretical expressions ($ 0.84 \sim C_{s} < C_{a} \sim 1.38$).
For both scalar and ALP fields, the radiation power is suppressed when $R_*  \gg C_{i}/\sqrt{1 - \xi^{2}}$.
The physical origin of this suppression is destructive interference between the emitted EM waves, which are emitted in phases from different locations within the particles.
It is remarkable that for the scalar case, one bump ($C_{s} \sim 3.43$) shows up in the radiated power besides the main radiation peak, whereas no such feature is observed in the ALP case.
This additional bump becomes particularly significant in the presence of the resonance effect, 
as shown in the left panel of Fig.~\ref{Fig: alt_mag_R}.

\begin{figure}[htbp]
    \begin{minipage}[b]{0.48\columnwidth}
        \centering
        \includegraphics[width=\columnwidth]{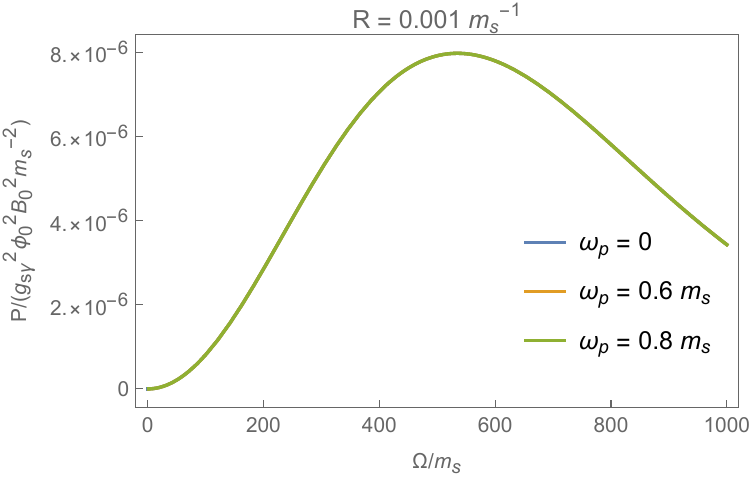}
    \end{minipage}
    \begin{minipage}[b]{0.48\columnwidth}
    \centering
    \includegraphics[width=\columnwidth]{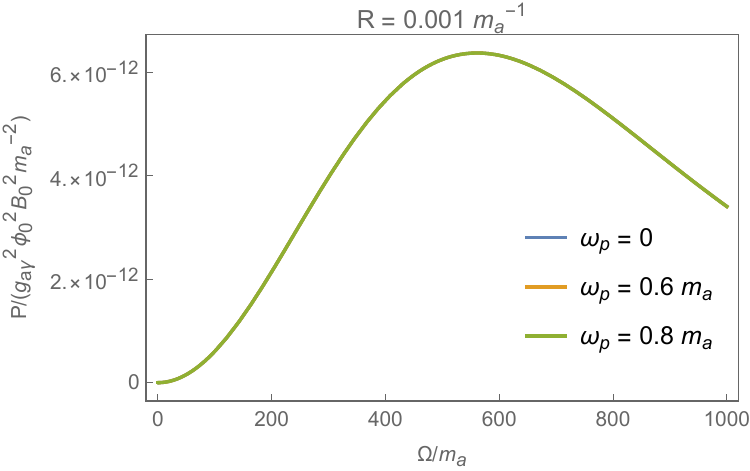}
    \end{minipage}
    \caption{
    The radiated power as a function of $\zeta = \Omega/m$ in the case of the alternating magnetic field with $R = 0.001 \omega^{-1}$
    for the scalar (Left panel) and ALP (Right panel). 
    } 
    \label{Fig: alt_mag_Omega}
\end{figure}

Next, we plot the radiated power in the alternating magnetic field in Figs.~\ref{Fig: alt_mag_R} and \ref{Fig: alt_mag_Omega}.
In Fig.~\ref{Fig: alt_mag_R}, we consider two types of resonance effects: 
$\omega \sim \Omega =0.999m $ and $\omega_{p} = 0$; 
$\omega \sim \omega_{p} =0.999m$ and $\Omega = 0$. 
As discussed in the previous two sections, the radiation peaks for scalar and ALP are characterized by $R_* \sim C_{i}/\sqrt{(1 \pm \zeta)^{2}-\xi^{2}}$.
We find that the resonance between the oscillating scalar/ALP field and the alternating magnetic field leads to more efficient radiation than the resonance between an oscillating field and a background plasma.
Moreover, the difference between the scalar and ALP cases becomes significant due to the resonance effect, which may provide further possibilities for distinguishing between these two types of fields through their radiation signatures.


In Fig.~\ref{Fig: alt_mag_Omega}, we consider three different values of the plasma frequency, $\omega_p/m = 0, 0.6, 0.8$, while fixing $R$ (equivalently, $R_*$) to $R = 0.001 m^{-1}$ ($R_*=R/m^{-1}=0.001$).
These two plots illustrate how the radiated power depends on the frequency $\Omega$ of the alternating magnetic field for different values of the plasma frequency $\omega_p$.
When $R \ll m^{-1}$, the resonance condition is approximately given as $\Omega \sim C_{i}/R \gg m, \omega_{p}$.
Thus, the plasma medium effect on the radiated power is negligible.
Consequently, the radiated powers for different choices of the plasma frequency $\omega_p$ overlap in Fig.~\ref{Fig: alt_mag_Omega}.

\begin{figure}[htbp]
    \begin{minipage}[b]{0.48\columnwidth}
        \centering
        \includegraphics[width=0.92\columnwidth]{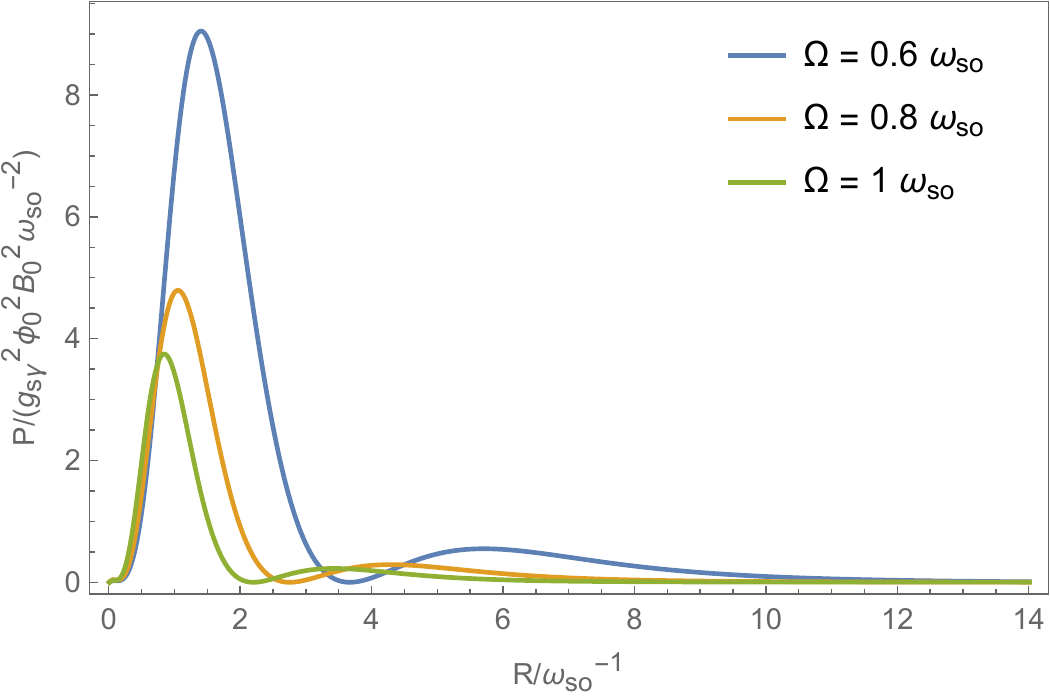}
    \end{minipage}
    \begin{minipage}[b]{0.48\columnwidth}
    \centering
    \includegraphics[width=\columnwidth]{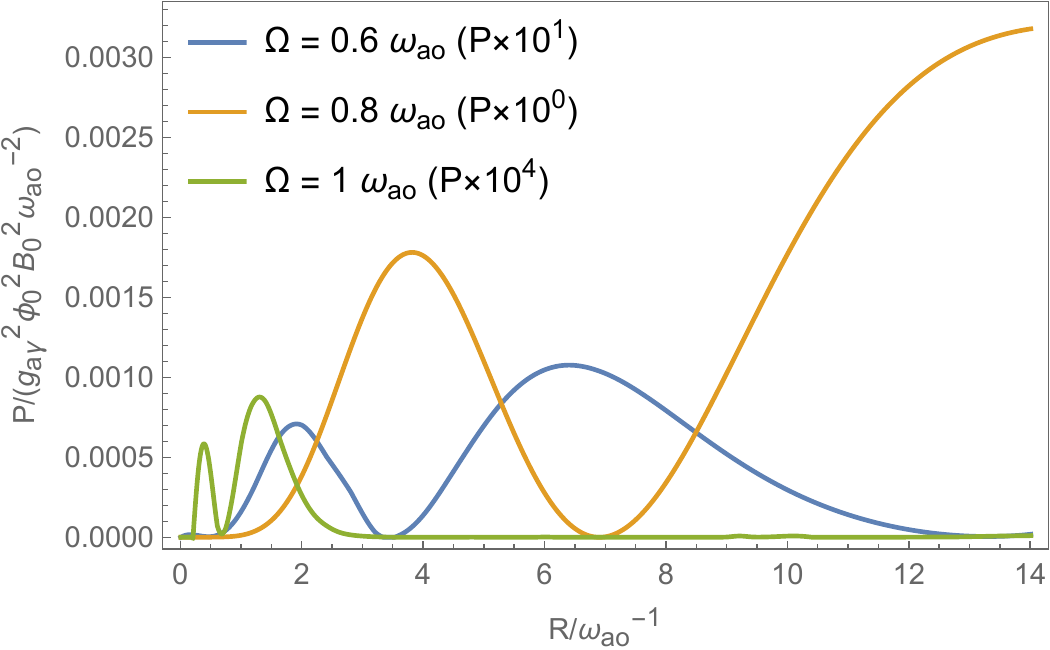}
    \end{minipage}
    \caption{
    The radiated power as a function of $R$ in the case of radial oscillation with $\delta_{R} =1/1000$
    for the scalar (Left panel) and ALP (Right panel).
    }
    \label{Fig4}
\end{figure}

Finally, we plot the radiated power in the radial oscillation of the scalar/ALP field in Fig.~\ref{Fig4}.
We have chosen the amplitude of radial oscillation $\delta_{R} =1/1000$ as a benchmark value, while considering three different values of the plasma frequency $\omega_p/m = 0, 0.6, 0.8$.
Remarkably, the radiation power for the scalar case is much larger than that for the ALP case.
Thus, the current setting provides a unique and concrete example highlighting how the different couplings to the EM field lead to observable differences in the resulting EM signatures.
Note that, to make an intuitive comparison between the radiation power behaviors of scalar and ALP cases, the radiation power values in the case of ALP for $\Omega = 0.6 \omega_{ao}$ and $\Omega = \omega_{ao}$ are multiplied by $10^{1}$ and $10^{4}$ respectively in Fig.~\ref{Fig4}.

Moreover, in the scalar-field case, the radiated power shows the same behavior as in the case with a constant magnetic field: the plasma medium effect is negligible, and no additional resonance structure arises. 
In contrast, in the case of the ALP field, a resonance effect different from the previous two cases shows up, and two characteristic peaks appear in the radiated power.

\subsection{Detectability}

It is of great significance that we analyze the detectability of the EM signals generated by scalar and ALP and the possibility of distinguishing them by observation.
We consider a mass range of $10^{-7}$ eV to $10^{-2}$ eV, corresponding to a frequency range of $24$ MHz to $2400$ GHz, which is detectable by existing and forthcoming radio telescopes.
Denoting the distance between the source and the Earth by $d$,
we obtain the flux of EM radiation reaching Earth as $F = L/(4 \pi d^{2})$
for the luminosity $L = \mathcal{P}$. 
The spectral flux density can be calculated as $S= F/\mathcal{B}$, 
where $ \mathcal{B} $ is the signal bandwidth.
We can take $\mathcal{B} = \triangle \nu_{\gamma} \sim \omega /2\pi$ and $\omega$ is the angular frequency of the EM signal~\footnote{
The bandwidth is estimated by the empirical relation in Fourier space $\Delta t \cdot \Delta \nu \gtrsim 1$, which leads to $\Delta \nu \sim 1/\Delta t \sim \omega/2\pi$.
}.

Then, the spectral flux density can be written as
\begin{align}
   S = \frac{L}{4 \pi d^{2} \mathcal{B} }
    =\frac{\mathcal{P}}{2 \omega d^{2}}
    \, .
\end{align} 
In particular, for the case of the alternating magnetic field, although the radiated power has contributions of two different frequencies $|m - \Omega|$ and $m + \Omega$, we consider that the detected spectral line has a frequency of $m /2 \pi$ when $m \sim \Omega$. 
\begin{figure*}[htbp]
    \centering
    \includegraphics[width=0.9\textwidth]{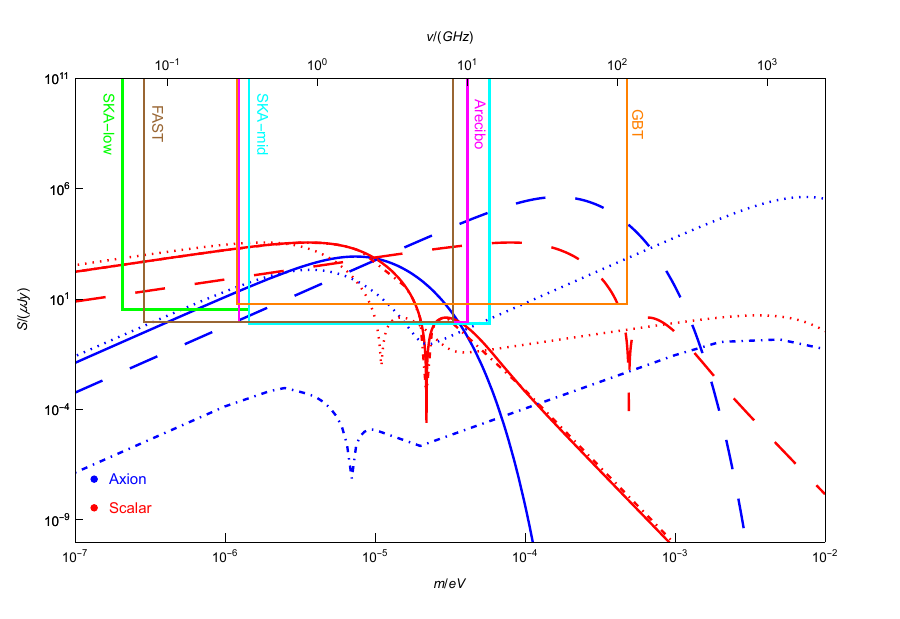}
    \caption{
    The blue and red curves represent the spectral flux densities of ALP and scalar, respectively.
    We plot spectral flux density in the four cases:
    the constant magnetic field which $\omega_{p} = 0$ (solid curves);
    the alternating magnetic field with $m \sim \omega_{p} = 0.999m $ and $\Omega =0$ (dashed curves), 
    and with $m \sim \Omega = 0.999m$ and $\omega_{p} =0$ (dotted curves);
    the radial oscillating with $\omega_{o} \sim \Omega = 0.999 m$ (dot-dashed curves).
    The minimum of detectable spectral flux density for the SKA-low, SKA-mid, FAST, Arecibo, and GBT are shown for $t_{{\rm obs}} = 1 {\rm hr}$~\cite{Bai:2017feq}.
    }
    \label{observation}
\end{figure*}

In Fig.~\ref{observation}, we plot four cases of spectral flux densities for different particle masses:
The constant magnetic field which $\omega_{p} = 0$; 
alternating magnetic field with $m \sim \omega_{p} $ and $m \sim \Omega $; 
radial oscillation with $\omega_{o} \sim \Omega$.
We also plot detection sensitivities for various telescopes:
Square Kilometre Array (SKA) can cover the frequency range of 50 MHz - 350 MHZ (SKA-low) and 350 MHz - 14 GHz (SKA-mid)~\cite{dewdney2009square};
Five hundred meter Aperture Spherical Telescope (FAST) can cover the range of 70 MHz - 8 GHz~\cite{Nan:2011um};
Arecibo Observatory (Arecibo) can cover the range of 300 MHz - 10 GHz~\cite{Giovanelli:2005ee}; 
Green Bank Telescope (GBT) can cover the range of 290  MHz - 115.3 GHz~\cite{prestage2009green, white2022green}.

In Fig.~\ref{observation}, the minimum of detectable spectral flux density for a radio telescope can be estimated as follows~\cite{Bai:2017feq}:
\begin{align}
    S_{min} 
    = 
    \text{SNR}_{min} \frac{\text{SEFD}}{\sqrt{n_{p} B t_{obs} }} 
    \, .
\end{align}
$\text{SNR}_{min}$ stands for the minimum signal-to-noise ratio, $n_{p}$ is the number of polarization, where we take $n_{p} = 1$, 
$B$ is the bandwidth, and $t_{obs}$ is the observation time.
$\text{SEFD} =2k_{B}T_{sys}/A_{eff}$ stands for the system-equivalent flux density, where $k_{B}$ is Boltzmann constant, $T_{sys}$ is the telescope system temperature, and $A_{eff}$ is the effective of telescope. 
$\text{SEFD}$ is frequency dependent, and we chose the typical values to estimate the minimum of the detectable spectral flux density.
Depending on the parameters of the different telescopes, we can estimate their minimum detectable spectral flux density~\cite{dewdney2009square, Nan:2011um, Giovanelli:2005ee, prestage2009green, Bai:2017feq, Hook:2018iia, white2022green}.

In astrophysical environments, such as the strong magnetic fields ($10^{8}$–$10^{15}$ G) surrounding neutron stars immersed in plasma, photons can acquire an effective mass approximately equal to that of a scalar or ALP~\cite{Amin:2021tnq, Buckley:2020fmh}.
Moreover, rotating neutron stars, orbiting binaries, and merging neutron stars can generate extremely strong time-varying magnetic fields~\cite{Sen:2021mhf, Kiuchi:2015sga}.
Note that neutron stars serve here merely as astrophysical laboratories rather than the objects of interest themselves.
Following Ref.~\cite{Bai:2017feq}, we assume $B_{0} = 10^{10} {\rm G}$ and $d = 1 {\rm kpc}$ as benchmark parameters.
For the ALP case, the relation between the characteristic size of the ALP star $R$ and mass $m_a$ is given as
\begin{align}
    R = 0.02 {\rm m} \times \left(\frac{4 \times 10^{12} {\rm GeV}}{f_{a}}\right)^{1/2} \left(\frac{10^{-5} {\rm eV}}{m_{a}}\right)^{1/2} \left(\frac{M_{a}}{10^{-16} M_{\odot}}\right)^{0.3}
    \, ,
\end{align}
where $f_{a}$ and $M_{a}$ are decay constant and ALP star mass respectively.
Considering the QCD axion, we fix the $m_{a} f_{a} = (2 \times 10^{8} {\rm eV})^{2}$~\cite{Bai:2017feq}. 
The coupling constant can be rewritten as $g_{a\gamma } = \alpha c_{\gamma}/(\pi f_{a})$, 
where $\alpha$ is the fine structure constant, 
$c_{\gamma}$ is a model-dependent number, and its value can vary from $\sim \mathcal{O}(1)$ to many orders of magnitude higher~\cite{Choi:2015fiu, Kaplan:2015fuy, Agrawal:2017cmd, Kouvaris:2022guf}.
For a dense ALP star, $\phi_{0} \sim {\cal O}(1) f_{a} $~\cite{Kyriazis:2022gvw}, 
and we fix the model parameters as $g_{a\gamma}\phi_{0} =10^{-2} $. 

We note that the term “ALP star” refers to a compact, gravitationally bound and coherent ALP condensate configuration rather than a stellar object in the conventional astrophysical sense. 
Its characteristic radius depends sensitively on the ALP mass, decay constant, self-interaction strength, and total mass of the condensate, and can vary over several orders of magnitude depending on the parameter choice. 
In the present analysis, the reference values adopted for numerical estimates correspond to the QCD axion relation 
$m_a f_a = (2 \times 10^8~\mathrm{eV})^2$, 
and should be regarded as representative benchmark parameters.

For the comparison of EM signal emission from the scalar with that from the ALP, we choose the same parameters for the scalar as those for the ALP, which allows us to focus on the different coupling to photons and their characteristic observational signatures.
As shown in Fig.~\ref{observation}, the resonance effect can enhance the EM radiation signal with respect to the value $m R$ as discussed in the previous subsection, and different resonance effects have different impacts on EM signals.
For the radial oscillation of the fields in alternating magnetic fields, the radiation signals generated by a scalar are in the detectable range. 
However, those by ALP are not enough to be detected even with the resonance effect.
Moreover, when we consider that $B_{0} = 10^{14}$G, the EM radiation from the scalar and ALP can account for the fast radio bursts~\cite{Petroff:2016tcr, Katz:2016dti}.

\section{
Implications for modified gravity
}

In our analysis so far, we have not specified the origin of the scalar field.
In this section, we revisit our formulation and parameter choices, assuming that the scalar field arises from a modified-gravity theory.

\subsection{
Scalar-tensor realization in modified gravity
}

As discussed in Sec.~II, the scalar field considered in this work can naturally arise in scalar–tensor theories of gravity, where the coupling to the electromagnetic field emerges through the Weyl transformation of the spacetime metric. 
Representative examples include Brans–Dicke theory~\cite{Brans:1961sx, Jordan1955} and its generalizations, as well as modified-gravity models such as $F(R)$ gravity~\cite{Nojiri:2010wj, Sotiriou:2008rp}, which can be reformulated as scalar–tensor theories in the Einstein frame. 
In the context of DE models, scalar–tensor realizations with runaway potentials, such as inverse power-law models, have been extensively studied (for review, see Ref.~\cite{observations:2022mik}).

We note that in a minimal $F(R)$ gravity, the coupling function takes the form $A(\phi)=\exp\left(-\sqrt{1/6}\,\phi/M_{\mathrm{pl}}\right)$,
so that the scalar–matter coupling strength is uniquely fixed by the gravitational sector and suppressed by the Planck scale. 
Expanding for small $\phi/M_{\mathrm{pl}}$, the scalar-photon coupling is therefore of the order of $1/M_{\mathrm{pl}}$, which is typically much smaller than the phenomenological range explored in this work. 
Our analysis should thus be regarded as a more general effective treatment beyond the minimal $F(R)$ gravity.
It is worth mentioning that our analysis can be applied to the $F(R)$ gravity coupled to the axion field, known as the axion $F(R)$ gravity~\cite{Odintsov:2020iui, Odintsov:2020nwm}.

\subsection{
Matter coupling and the chameleon mechanism
}

First, using the chameleon mechanism unique to modified gravity, we revisit the matter coupling that appeared in Sec.~3.
We recall the coupling $g_{s \gamma} \phi J_{0}^{\mu}$ in Eq.~\eqref{Eq: scalar_matter_coupling} in Sec.~III.
This coupling generally affects physical constants, such as the fine-structure constant and masses of the fundamental particles. 
Such variations of physical constants originate from the scalar-mediated interaction, and the chameleon mechanism can provide a way to suppress the large variations in physical constants~\cite{Brax:2013ida, Burrage:2017qrf, Brax:2021wcv}.

The chameleon mechanism is triggered by the ambient matter distribution (the background current $J_0^\mu$ in our case) and increases the mass of the scalar field.
In other words, the coupling between the scalar field and background current can be understood in light of the chameleon mechanism.
We can discuss EM radiation in the context of the background configuration, accounting for the chameleon mechanism and the effective mass.

\subsection{Synergy with ground-based experiments}

Finally, we consider the range of the scalar-field mass.
For the viable DE models of the scalar-tensor theory with the chameleon mechanism, it is not surprising that the effective mass is larger than the DE scale.
And, it is plausible to consider the mass range as in Fig.~\ref{observation} in the framework of the scalar-tensor theory.
For instance, existing and planned experiments searching for the DE scalar field scan the mass range $10^{-7}~\mathrm{eV} - 10^{-2}~\mathrm{eV}$~\cite{Burrage:2017qrf, Katsuragawa:2021wmw}.
Also, our parameter choice for the coupling constant and field oscillation amplitude adopted in the previous section, $g_{s\gamma}\phi_0 = 10^{-2}$, can be consistently embedded in the scalar-tensor theory.

\begin{figure*}[htbp]
    \centering
    \includegraphics[width=0.9\textwidth]{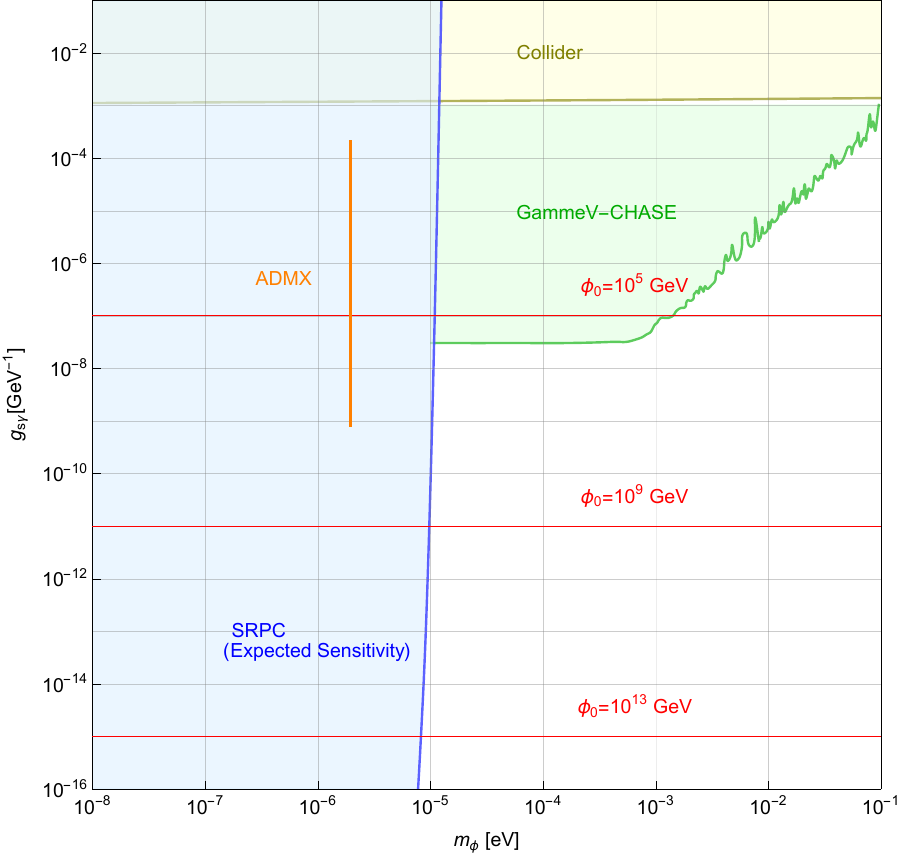}
    \caption{
    The red horizontal lines correspond to the $g_{s\gamma}\phi_0 = 10^{-2}$ with different choices of $\phi_0$. 
    We show existing constraints from GammeV Chameleon
    Afterglow Search (GammeV-CHASE) by green region~\cite{Steffen:2010ze}, Axion Dark Matter Experiment (ADMX) by orange line~\cite{ADMX:2010ndb}, and collider experiments by yellow region~\cite{Brax:2010gp}, as well as the expected sensitivity of Stimulated Resonant Photon Collider (SRPC) by blue region~\cite{Homma:2022ktv}.
    }
    \label{sensitivity-curve}
\end{figure*}

We plot the existing and planned experiments searching for the DE scalar field in Fig.~\ref{sensitivity-curve}.
Here, fixing $g_{s\gamma}\phi_0 = 10^{-2}$, we plot red lines corresponding to given field amplitude $\phi_0$.
As in this figure, the EM signal from the scalar field can surely constrain the modified gravity in synergy with the ground-based experiments,
which potentially provides the new observational tests of the modified gravity scenario alongside ALP models.
Note that the field amplitude $\phi_0$ is the model parameter included in the ansatz for the scalar/ALP condensate in our analysis.
However, it is possible to determine the field amplitude by fully solving the Klein-Gordon equation of the scalar/ALP field, and we can quantitatively estimate the scalar-field mass in a concrete model of scalar-tensor theories and background matter distribution.
In other words, it is also possible to place bounds on the model parameters in the scalar-tensor theory using astrophysical EM signals, as we have observational bounds on the mass and coupling constants for axion and ALPs.

\section{Conclusion and discussion}

We have investigated the EM radiations generated by an oscillating scalar field, comparing them with those generated by the ALP field.
While our analytical methods are adapted from existing studies of axion and ALP-induced EM radiation, we have demonstrated that their application to pure scalar fields reveals both qualitative and quantitative differences in radiation behavior. 
This approach serves not only as a theoretical extension of existing methods but also as a complementary framework for exploring observational avenues to test scalar degrees of freedom in modified gravity.

We have shown that two different resonance effects can significantly enhance the EM radiation in specific regimes,
and that the nature of these enhancements depends sensitively on the form of the coupling, $F_{\mu\nu} \tilde{F}^{\mu\nu}$ for ALPs and $F_{\mu\nu} F^{\mu\nu}$ for pure scalars. 
Our comparative analysis of constant and alternating magnetic field backgrounds reveals that scalar fields often exhibit less efficient radiation than ALPs, especially in the regime where the scalar Compton wavelength is smaller than the system size. 
Notably, in radially oscillating configurations, the ALP field can produce a much stronger EM signal than the scalar field, highlighting how field configuration also plays a critical role in detectability.

We have found that in a constant magnetic field, both scalar and ALP fields produce negligible EM radiation when $m R \gg 1$ unless the frequency of the background plasma is close to that of the oscillating field.
We have observed that in a background alternating magnetic field, similar radiation behavior occurs, and the resonance effect depends on the plasma frequency and the oscillating scalar/ALP frequency.
The radiation enhancements by resonance effects are significant enough to clarify the difference between scalar and ALP.

We have also considered the radial oscillation of the scalar and ALP fields in an alternating magnetic field background. 
The resonance effect can enormously enhance the radiated power generated by the ALP field.
In the specific range $m R$, the EM signal generated by the ALP condensate is much stronger than the EM signal generated by the scalar field.
We have found that the magnitude of the radial oscillation has a more significant effect on the value of the radiated power generated by the ALP than the scalar.

Although we have focused on specific field configurations, our formulation is general and can be extended to a broad class of scalar and ALP sources. 
We also emphasized the importance of incorporating astrophysical parameters such as ambient plasma frequency and magnetic field strength, which influence the observability of the radiation. 
As in Eq.~\eqref{Eq: scalar_matter_coupling}, the inclusion of matter coupling offers another potentially observable distinction, and the chameleon-mechanism effects, unique to the scalar field originating from modified gravity, may arise.
It is also essential to investigate the stability of the scalar/ALP configurations,
providing the time scale of the EM signals.
These dense stars can be collectively discussed as boson stars~\cite {Schunck:2003kk}, and understanding their dynamical evolution, including the backreaction from EM radiation, will be important.

Moreover, we have focused on EM signals generated by the single scalar/ALP star.
Future work should address signal event rates in realistic astrophysical contexts. 
As discussed in Refs.~\cite{Bai:2017feq, Buckley:2020fmh, Amin:2021tnq}, we can examine the encounter rate of the scalar/ALP stars in a strong magnetic field source, which can be converted into the event rate.
At the same time, it is also significant to evaluate the distinguishability between the EM signals from scalar/ALP stars and other astrophysical signals.
A careful comparison with other astrophysical emission mechanisms will be essential to assess the feasibility of detection. 
We shall address those in future work.

We make several final remarks.
The qualitative analysis of the radiation power in this work is valid for any mass scale of the scalar and ALP field. 
In our analysis of the detectability of scalar and ALP, we have chosen the mass range $10^{-7} - 10^{-2}$ eV, which contains the ALP mass scales that can explain the DM density in the Universe~\cite{Turner:1989vc,cirelli2024dark}.
Our model and analytic calculations can be applied to any pure or pseudo-scalar field, and the scalar and ALP mass scales are extensive. 
Considering the practical scenarios and fundamental theories, we can perform detailed studies and investigate observational predictions.

Regarding the scalar-field case, we have observed the unique matter coupling as in Eq.~\eqref{Eq: scalar_matter_coupling}, which is absent in the ALP case.
It is thus intriguing to investigate the effect of the matter coupling and take it into our current analysis.
In light of the modified gravity as an origin of the scalar field, the mass range can be computed by specifying the ambient matter field.
Setting the actual astrophysical environment, such as the magnetosphere around neutron stars or black holes, to analyze the chameleon mechanism allows for more precise calculations.
In parallel with axion and ALP searches, it would be essential to explore the new aspects of astroparticle physics related to the fundamental scalar field in modified gravity theories.
Ultimately, our work provides a theoretical foundation for distinguishing pure and pseudo-scalar fields based on their EM signatures and highlights new opportunities in astroparticle physics beyond the standard ALP paradigm.


\section*{Acknowledgement}

T.K. is supported by National Key R\&D Program of China (No.~2021YFA0718500), National Natural Science Foundation of China (No.~f), and Natural Science Foundation of Hubei Province (No.~2022CFB817). 
S.N. is supported by JSPS KAKENHI Grant No.~24K17053.
T.K. thanks Shinya Matsuzaki for his fruitful comments.
W.W. thanks the astrophysics group at Central China Normal University for valuable discussions.


\appendix
\section{Calculation appendix}

\subsection{Time-averaged radiation power in scalar case}

In the calculation of Eq.~\eqref{Eq: pta_scalar}, we used the following result for the basis vectors:
\begin{align}
\begin{split}
    &\hat{\mathbf{x}} \cdot 
    \left \{ 
        (\hat{\mathbf{x}}\times \hat{\mathbf{E}}_{z}) 
        \times 
        \left[
            \hat{\mathbf{x}} \times(\hat{\mathbf{x}}\times \hat{\mathbf{E}}_{z})
        \right]
    \right \}
    \\
    &= 
    \hat{\mathbf{x}} \cdot \{ \hat{\mathbf{x}}(\hat{\mathbf{x}} \times \hat{\mathbf{E}}_{z}) \cdot (\hat{\mathbf{x}} \times \hat{\mathbf{E}}_{z})- (\hat{\mathbf{x}} \times \hat{\mathbf{E}}_{z})[(\hat{\mathbf{x}} \times \hat{\mathbf{E}}_{z}) \cdot \hat{\mathbf{x}} ]\}
    \\
    &= 
    \hat{\mathbf{x}} \cdot (\hat{\mathbf{x}} |\hat{\bm{\varphi}}| |\hat{\bm{\varphi}}| )
    \\
    &=
    |\hat{\mathbf{x}}|^{2} |\hat{\bm{\varphi}}|^{2}
    \\
    &= 1 
    \, .
\end{split}
\end{align}
Here, we denote the unit vector in the azimuth direction by $\hat{\bm{\varphi}}$.
In the first line of the above equation, $\hat{\mathbf{x}}\times \hat{\mathbf{E}}_{z}$ and $\hat{\mathbf{x}} \times(\hat{\mathbf{x}}\times \hat{\mathbf{E}}_{z})$ corresponds to the basis of $\mathbf{E}_{r}\left(\mathbf{x}, t\right)$ and $\mathbf{B}_{r}\left(\mathbf{x}, t\right)$ via Eq.~\eqref{Eq: EMfield}.
Their cross product represents the basis of the Poynting flud ${\bf S} \left(\mathbf{x}, t\right)$,
and the inner product with $\hat{\mathbf{x}}$ appears in Eq,~\eqref{Eq: power_per_unit}.
Note that $\nabla A_{0}$ does not contribute to the radiated power because of the absence of the charge density $\rho_{s}\left (\mathbf{x}, t\right)=0$.
The charge density vanishes because the background electric field is assumed to be zero.

\subsection{Time-averaged radiation power in ALP case}

Compared with the scalar field case, we find that the different coupling to the EM field results in different basis vectors in the current ${\bf J}_{s}$.
In the calculation of Eq.~\eqref{Eq: pta_axion}, we used the following result for the basis vectors:
\begin{align}
\begin{split}
    &\hat{\mathbf{x}} \cdot [\hat{\mathbf{E}}_{z}\times (\hat{\mathbf{x}}\times \hat{\mathbf{E}}_{z})]
    \\
    &= \hat{\mathbf{x}} \cdot [\hat{\mathbf{x}} |\hat{\mathbf{E}}_{z}|^{2} - \hat{\mathbf{E}}_{z} (|\hat{\mathbf{x}}||\hat{\mathbf{E}}_{z}|\cos \theta)]
    \\
    &= |\hat{\mathbf{x}}|^{2} |\hat{\mathbf{E}}_{z}|^{2} - |\hat{\mathbf{x}}|^{2} |\hat{\mathbf{E}}_{z}|^{2} \cos \theta^{2}
    \\
    &= \sin^{2}\theta
    \, .  
\end{split}
\end{align}
In the first line of the above equation, $\hat{\mathbf{E}}_{z}$ and $\hat{\mathbf{x}} \times \hat{\mathbf{E}}_{z}$ corresponds to the basis of $\mathbf{E}_{r}\left(\mathbf{x}, t\right)$ and $\mathbf{B}_{r}\left(\mathbf{x}, t\right)$ via Eq.~\eqref{Eq: EMfield}.
Their cross product represents the basis of the Poynting flud ${\bf S} \left(\mathbf{x}, t\right)$,
and the inner product with $\hat{\mathbf{x}}$ appears in Eq.~\eqref{Eq: power_per_unit}.
We note that $\nabla \mathbf{A}_{0}$ does not contribute to the radiated power, as in the scalar field case, but for a different reason.
The charge density does not vanish because the background magnetic field is nonzero.
In calculating $dP/d\Omega$, $(\mathbf{E}_{r} \times \mathbf{B}_{r}) \cdot \hat{\mathbf{x}} $ includes $\nabla \mathbf{A}_{0}$, 
however, it is proportional to $\left[\hat{\mathbf{x}} \times \left(\hat{\mathbf{x}} \times \hat{\mathbf{E}}_{z}\right)\right] \cdot \hat{\mathbf{x}}= 0$.


\bibliographystyle{unsrt}
\bibliography{References}

\end{document}